\documentclass[aps,prc,preprint,superscriptaddress,nofootinbib,nopreprintnumbers,floatfix,noshowpacs,noshowkeys,a4paper,titlepage]{revtex4-1}

\usepackage[pdftex]{graphicx}% Include figure files
\usepackage{color}

\usepackage{hyperref}

\usepackage[centertags]{amsmath}
\usepackage{amssymb}
\usepackage{slashed}

\usepackage{microtype} %Nice typing

\def\be{\begin{equation}}
\def\ee{\end{equation}}
\def\bea{\begin{eqnarray}}
\def\eea{\end{eqnarray}}
\def\bear{\begin{array}}
\def\ear{\end{array}}
\def\bfig{\begin{figure}}
\def\efig{\end{figure}}
\def\bcen{\begin{center}}
\def\ecen{\end{center}}
\def\bi{\begin{itemize}}
\def\ei{\end{itemize}}

\def\raw{\rightarrow}

\def\slash{\!\!\! /}
\begin{document}
\title{Charged kaon production by coherent scattering of neutrinos and antineutrinos on nuclei
% Kaon Production Via Coherent Scattering of Neutrinos \\
%       or Kaon production by coherent scattering of neutrinos on nuclei \\
%       or Coherent kaon production in neutrino-nucleus scattering
%       or Coherent kaon production induced by neutrinos
}

\author{L. \surname{Alvarez-Ruso}}
\affiliation{Centro de F\'\i sica Computacional, Departamento de F\'\i sica, Universidade de Coimbra, Portugal}
\affiliation{Instituto de F\'isica Corpuscular (IFIC), Centro Mixto
Universidad de Valencia-CSIC, E-46071 Valencia, Spain}
\author{J. \surname{Nieves}}
\affiliation{Instituto de F\'isica Corpuscular (IFIC), Centro Mixto
Universidad de Valencia-CSIC, E-46071 Valencia, Spain}
\author{I. \surname{Ruiz Simo}}
\affiliation{Departamento de F\'\i sica Te\'orica and IFIC, Centro Mixto
Universidad de Valencia-CSIC, E-46071 Valencia, Spain}
\affiliation{Departamento de F\'isica At\'omica Molecular y Nuclear, Universidad de Granada, E-18071 Granada, Spain}
\author{M. \surname{Valverde}}
\affiliation{Research Center for Nuclear Physics (RCNP), Osaka University, Ibaraki, 567-0047, Japan}
\author{M. J.  \surname{Vicente Vacas}}
\affiliation{Departamento de F\'\i sica Te\'orica and IFIC, Centro Mixto
Universidad de Valencia-CSIC, E-46071 Valencia, Spain}

\begin{abstract}
With the aim of achieving a better and more complete understanding of neutrino interactions with nuclear targets, 
the coherent production of charged kaons induced by neutrinos and antineutrinos is investigated in the energy range of some of the current neutrino experiments. We follow a  microscopic approach which, at the nucleon level, incorporates the most important mechanisms allowed by the chiral symmetry breaking pattern of QCD. The distortion of the outgoing $K$ ($\bar{K}$) is taken into account by solving the Klein-Gordon equation with realistic optical potentials.  Angular and momentum distributions are studied, as well as the energy and nuclear dependence of the total cross section. 
\end{abstract}

\date{\today}


\maketitle

\section{Introduction}

In the new era of precise neutrino oscillation experiments, a good understanding of neutrino scattering cross sections are crucial to have a realistic simulation of the detection process and reduce systematic errors that will be soon taking over the statistical ones. Research on these cross sections from both theoretical and experimental sides are also relevant for hadronic and nuclear physics as they enlarge the information on hadronic and nuclear structure complementary to the one obtained with other probes.

In the few-GeV region, the attention has been focused on the processes with the largest cross sections (quasielastic and pion emission) but strangeness production is also relevant. For example, the $\nu_l \, N \raw l^- \, K^+ \, N'$  process induced by atmospheric neutrinos is a background for one of the candidates for hypothetical proton decay mechanisms ($p \raw \bar{\nu} \, K^+$), when the final lepton escapes detection~\cite{Kobayashi:2005pe,MarrodanUndagoitia:2006qn}. A better understanding of antikaon ($\bar{K}$) production is important for experiments that will take data in the $\bar{\nu}$ mode such as MINER$\nu$A, NO$\nu$A and T2K. In this regime, single hyperon production measurements allow to extract transition form factors and Cabibbo-Kobayashi-Maskawa matrix elements~\cite{Solomey:2005rs}. In addition these hyperons can decay inside the detectors and  contribute to the pion yield at low energies~\cite{Singh:2006xp}.

Neutrino-induced strange-particle production cross sections are poorly known.  After the first bubble chamber events with positive kaons and hyperons~ \cite{Barish:1974ye}, few other results have been reported~\cite{Barish:1978pj,Baker:1981tx}. Moreover, no such measurements exist with  $\bar{\nu}$ fluxes. One should recall that associated strangeness production ($\Delta S =0$) has a high threshold  because both a kaon and a hyperon are emitted; instead, single $K$, hyperon ($\Delta S =-1$) and $\bar{K}$  ($\Delta S =1$) production are Cabibbo suppressed. The experimental situation shall improve in the near future thanks to the MINER$\nu$A experiment which will allow for high-statistics studies of exclusive strangeness production reactions~\cite{Solomey:2005rs}.

On the theoretical side, after the pioneering papers of Refs.~\cite{Shrock:1975an,Mecklenburg:1976pk,Amer:1977fy,Dewan:1981ab}, addressing associated strangeness~\cite{Shrock:1975an,Mecklenburg:1976pk,Amer:1977fy}, single hyperon production~\cite{Amer:1977fy} and other $\Delta S = \pm 1$ reactions~\cite{Dewan:1981ab}, new work has emerged only recently~\cite{Singh:2006xp,Mintz:2006yp,Mintz:2007zz,Adera:2010zz,Adera:2011wr,RafiAlam:2010kf,Alam:2012zz}. References.~\cite{Singh:2006xp,Mintz:2006yp,Mintz:2007zz} use SU(3) symmetry and phenomenological information about nucleon form factor and hyperon decays to calculate the cross sections for  $\bar{\nu}_l \, N \raw l^+ \, Y$, with $Y = \Lambda,\,\Sigma$. A similar study was performed by Adera et al.~\cite{Adera:2010zz,Adera:2011wr} for charge-changing associated strangeness production $\nu_l \, N \raw l^- \, K \, Y$ in the threshold region. Finally, a model for  $\Delta S = \pm 1$ single (anti)kaon production processes $\nu_l \, N \raw l^- \, K \, N'$ and  $\bar{\nu}_l \, N \raw l^+ \, \bar{K} \, N'$ close to threshold based on SU(3) chiral Lagrangians was developed in Refs.~\cite{RafiAlam:2010kf,Alam:2012zz}. It has been stressed that the Monte Carlo generators employed in the analysis of neutrino experiments are not well suited to describe strangeness production at low energies and often underestimate the cross sections~\cite{RafiAlam:2010kf}.

With the exception of Refs.~\cite{Singh:2006xp,Adera:2011wr}, all the theoretical studies mentioned above assume single nucleon targets. However, all neutrino experiments are performed on nuclear targets, for which nuclear medium effects and final state interactions of the outgoing particles play an important role. One of the possible reaction channels that occur for nuclear targets is the coherent one, where the nucleus remains in the ground state. In the case of weak strangeness production, coherent reactions are possible for single charged $K^{\pm}$ production, namely 
\begin{equation}
  \nu_l (k) +\, {}^{A}\! Z_{\text{gs}}(p_A)  \to l^- (k^\prime) + \,
  {}^{A}\! Z_\text{gs}(p^\prime_A) +\, K^+(p_K) \, ,
\label{eq:reac1}
\end{equation}
and
\begin{equation}
  \bar{\nu}_l (k) +\, {}^{A}\! Z_{\text{gs}}(p_A)  \to l^+ (k^\prime) + \,
  {}^{A}\! Z_\text{gs}(p^\prime_A) +\, K^-(p_K) \,.
\label{eq:reac2}
\end{equation}

The coherent production of pions induced by neutrinos has received special attention as a potential 
background that may limit the sensitivity of neutrino oscillation measurements. In particular, 
neutral current coherent $\pi^0$ production   ($\nu \, {}^{A}\! Z_\text{gs}  \raw \nu \, \pi^0 \,  {}^{A}\! Z_\text{gs}$) 
is crucial for $\nu_e$ appearance searches: when one of the two photons from a $\pi^0$  decay is not detected, the $\pi^0$ 
cannot be distinguished from an electron born in a $\nu_e$ charged current interaction. Although charged current coherent $\pi^+$ 
production ($\nu_l \, {}^{A}\! Z_\text{gs}  \raw l^- \, \pi^+ \,  {}^{A}\! Z_\text{gs}$)  has been measured in the past at high energies, modern experiments K2K and SciBooNE could only obtain upper bounds at $E_\nu \sim 1$~GeV, in disagreement with their Monte Carlo simulations~\cite{Hasegawa:2005td,Hiraide:2008eu}. This unexpected result triggered a renewed theoretical interest in this process~\cite{Paschos:2005km,Singh:2006bm,AlvarezRuso:2007tt,Berger:2008xs,Amaro:2008hd,Hernandez:2010jf,Leitner:2009ph,Martini:2009uj,Nakamura:2009iq}. A recent short review of the present status with emphasis on the theoretical models can be found in Ref.~\cite{AlvarezRuso:2011zz}. In brief, coherent pion production models can be classified as PCAC and microscopic. PCAC models~\cite{Paschos:2005km,Berger:2008xs} use the partial conservation of the axial current (PCAC) to relate neutrino-induced coherent pion production to pion-nucleus elastic scattering. This simple and elegant description has some drawbacks at $E_\nu < 2$~GeV~\cite{Hernandez:2009vm}. Microscopic approaches~\cite{Singh:2006bm,AlvarezRuso:2007tt,Amaro:2008hd,Nakamura:2009iq} rely on models for pion production on the nucleon (performing a coherent sum over all nucleonic currents), implement nuclear effects and take into account the distortion of the outgoing pion wave. Their validity is restricted to the kinematic region where the pion production and distortion models are applicable.

Inspired by the theoretical developments outlined above on single kaon production and coherent pion production, we have investigated the coherent production of charged kaons induced by (anti)neutrinos [Eqs.~(\ref{eq:reac1},\ref{eq:reac2})] at low energies within a microscopic approach that follows Refs.~\cite{AlvarezRuso:2007tt,Amaro:2008hd}.  We implement the kaon production models on the nucleon of Refs.~\cite{RafiAlam:2010kf,Alam:2012zz} and account for the distortion of outgoing mesons using realistic descriptions of the (very different) interaction of $K$ and $\bar{K}$ in the nuclear medium.       
In Secs.~\ref{sec:formnu},\ref{sec:formanu} we briefly describe the formalism for kaon and antikaon production on the nucleon developed in Refs.~\cite{RafiAlam:2010kf,Alam:2012zz}, present the model for the coherent reaction and for the distortion of the outgoing kaons. Results are shown and discussed in  Sec.~\ref{sec:res}, to conclude with a summary in Sec.~\ref{sec:sum}.

\section{Formalism for $K^+$ coherent production}
\label{sec:formnu}

\subsection{Single kaon production model}
\label{subsec:micronu}

\begin{figure}[h!]
\begin{center}
\includegraphics[width=0.9\textwidth]{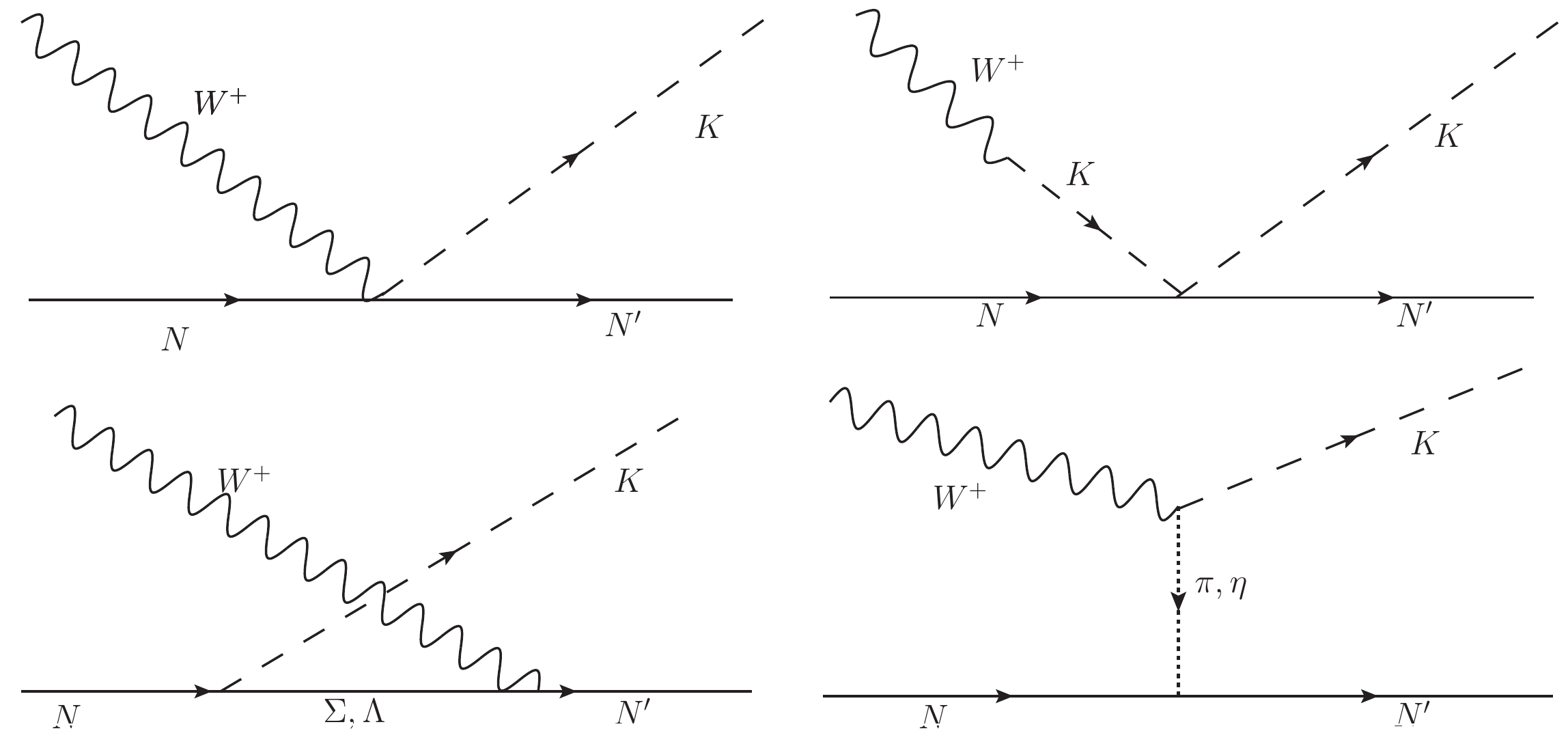}
\caption{\label{fig:diags}
  Feynman diagrams for $W^+ N\to  N K$. From the
  upper left corner in clockwise order: contact term (CT), Kaon pole term (KP),
  $\pi$ and $\eta$ in flight ($\pi$P, $\eta$P) and u-channel hyperon exchange (Cr$\Sigma$, Cr$\Lambda$) terms.} 
\end{center}
\end{figure}

For the elementary process $\nu_{l} \, p(n) \rightarrow l^- \, K^+ \, p(n)$ we adopt the description of Ref.~\cite{RafiAlam:2010kf} where the reaction mechanisms are derived from a Lagrangian that implements the QCD chiral symmetry breaking pattern. Although the vertices are SU(3) symmetric, this flavor symmetry is broken in the amplitudes by the physical hadron masses. This yields the set of diagrams for the hadronic currents shown in Fig.~\ref{fig:diags}, labeled as contact (CT), kaon pole (KP), u-channel crossed $\Sigma$
(Cr$\Sigma$) and $\Lambda$ (Cr$\Lambda$), pion in
flight ($\pi$P) and eta in flight ($\eta$P) terms. Owing to the absence of $S=1$ baryons, there are no s-channel amplitudes with $\Lambda$ or $\Sigma$ in the intermediate state.  
The structure of these currents is dictated by chiral symmetry with the couplings fixed from pion decay, nucleon and hyperon semileptonic decays and measured values of nucleon magnetic moments~\cite{RafiAlam:2010kf}. PCAC is implemented for the axial part of the currents. 
As the dependence of the different terms of the hadronic current on the momentum transferred to the nucleon is poorly known, if at all, the authors of Ref.~\cite{RafiAlam:2010kf} adopted a global dipole form factor $F(q^2)=(1-q^2/M_F^2)^{-2}$, with $M_F=1$~GeV  [$q^2=(k-k')^2$]. The cross section sensitivity to the variation of $M_F$ is studied in Ref.~\cite{RafiAlam:2010kf}, but will not be considered here once we regard the nucleon model as an input, with parameters extracted elsewhere. In the validity region assumed for the model ($E_\nu \leq 2$~GeV~\cite{RafiAlam:2010kf}), CT is the dominant contribution and interferes destructively with the rest. 

\subsection{The coherent reaction}
\label{subsec:cohenu}

The unpolarized differential cross section for reaction~(\ref{eq:reac1}) in the Laboratory frame can be cast as
\begin{equation}
\label{eq:sec}
\frac{d^{\,5}\sigma}{d\Omega_l dk_0' d\Omega_K } = \frac{1}{4 (2 \pi)^5}  
\frac{|\vec{k}^\prime||\vec{p}_K|}{|\vec{k}| M^2}\frac{G^2}{2}
L_{\mu\nu}\, {\cal A}^\mu_{K^+}(q,p_K) \left({\cal A}^\nu_{K^+}(q,p_K)\right)^*
\end{equation}
with $G$ and $M$ the Fermi constant and nucleon mass respectively. The leptonic tensor is
\begin{equation}
\label{eq:lep}
L_{\mu\nu}=
 k^\prime_\mu k_\nu +k^\prime_\nu k_\mu
- g_{\mu\nu} k\cdot k^\prime + {\rm i}
\epsilon_{\mu\nu\alpha\beta}k^{\prime\alpha}k^\beta \,,
\end{equation}
with $\epsilon_{0123}= +1$. The nuclear current ${\cal A}^{\mu}_{K^+}$ is obtained as the coherent sum over all nucleons, leading to the nuclear densities\footnote{Proton and neutron matter densities, normalized to the number of protons and neutrons in the nucleus, are taken from electron scattering data~\cite{DeJager:1974dg} and Hartree-Fock calculations~\cite {Negele:1972zp}, respectively~\cite{Nieves:1993ev}. They have been deconvoluted to get center point densities following the procedure described in Ref.~\cite{Oset:1989ey}.}  
\begin{equation}
{\cal A}^\mu_{K^+}(q,p_K) = 
\int d^3\vec{r}\ e^{{\rm i} \vec{q}\cdot\vec{r}}
\left\{\rho_p(\vec{r}\,) {\cal J}^\mu_{pK^+}(q,\hat{p}_K) + 
       \rho_n(\vec{r}\,) {\cal J}^\mu_{nK^+}(q,\hat{p}_K) \right \} \phi^*_{>}(\vec{p}_K,\vec{r})
\label{eq:A}
\end{equation}
where
\begin{equation}
{\cal{J}}^\mu_{NK^+}(q,\hat{p}_K) = \frac12 \sum_i \mathrm{Tr} \left[ (p\slash + M)\gamma^0 \Gamma_{i; NK^+}^\mu(q,\hat{p}_K) \right] \frac{M}{p_0} \,.
\label{eq:J}
\end{equation}
Index $i$ refers to all the possible mechanisms in Fig.~\ref{fig:diags}; $\Gamma_{i; NK^+}^\mu$ can be directly read from Eq.~(15) of Ref.~\cite{RafiAlam:2010kf} following the notation $j^\mu_i = \bar{N}(p') \Gamma_{i; NK^+}^\mu  N(p)$. 
To derive Eq.~(\ref{eq:J}), the initial and final nucleons in the nucleus, whose momenta are not well defined, are assumed to be on shell with $\vec{p} = (\vec{p}_K - \vec{q})/2$ and $\vec{p}\,' = - \vec{p}$. In this way the momentum transferred to the nucleus is equally shared by the initial and final nucleons. This approximation, which allows for a consistent description of the pion-nucleon and pion-nucleus kinematics, is based on the fact that, for Gaussian nuclear wave functions, it leads to an exact treatment of the terms linear in momentum of the elementary amplitude. More details can be found in the discussion between Eqs.~(7) and (8) of Ref.~\cite{Amaro:2008hd} and in references therein.

%\subsection{Kaon Distortion}
%\label{subsec:distort}

In Eq.~(\ref{eq:sec}), $\phi^*_{>}(\vec{p}_K,\vec{r})$ denotes the outgoing kaon wave function which we obtain as the solution of the Klein-Gordon equation 
\be
\left( - \vec{\nabla}^2 - \vec{p}_{K}^{\,2} + 2 p_K^0 V_{\mathrm{opt}} \right) \phi^*_{>}(\vec{p}_K,\vec{r}) =0 \,.
\label{eq:kg}
\ee 
The distorted wave Born approximation adopted here implies that the kaon momenta in $\Gamma_{i; NK^+}^\mu(q,\hat{p}_K)$ should be understood as operators acting on $\phi^*_{>}$: $\hat{p}_K \phi^*_{>} = (p_K^0 \phi^*_{>}, i\, \vec{\nabla} \phi^*_{>})$. This nonlocal treatment of kaon momenta affects only the (Cr$\Sigma$) and (Cr$\Lambda$) mechanisms. 

The optical potential $V_{\mathrm{opt}}$ characterizes the kaon interaction with the nuclear medium and is related to the in-medium kaon selfenergy $\Pi = 2p_K^0 V_{\mathrm{opt}}$. $\Pi$ is smooth at low energies due to the absence of $S=1$ baryon resonances and well described by the low density limit or $t \,\rho$ approximation, where $t$ is the forward kaon-nucleon elastic scattering amplitude. The real part of $\Pi$ is repulsive and, in a chiral SU(3) approach, dominated by the Weinberg-Tomozawa term~\cite{Waas:1996fy}. As the energy increases from threshold, the imaginary part of $\Pi$ coming from quasielastic, charge exchange $K^+ \,n \raw K^0 \, p$  and pion production $K \, N \raw K' \, N' \, \pi$ becomes sizable. It can be estimated by relating $\mathrm{Im}(t)$ to the kaon-nucleon total cross section $\sigma_{\mathrm{tot}}$ via the optical theorem, keeping in mind that this procedure might lead to some overestimation of   $\mathrm{Im}(\Pi)$ at low kaon energies because Pauli blocking and other in-medium corrections are neglected. Altogether
\be
2 p_K^0 V_{\mathrm{opt}} = \Pi = C m_K^2 \frac{\rho}{\rho_0} - i \, |\vec{p}_K| \sum_{N=p,n} \rho_N \sigma_{\mathrm{tot}}^{(K^+N)} \,.
\label{eq:opt}
\ee  
Here, $C=0.13$~\cite{Oset:2000eg,Cabrera:2004kt}, $\rho = \rho_p + \rho_n$ and $\rho_0 = 0.17$~fm$^{-3}$ is the normal nuclear density; $\vec{p}_K$ is taken in the Laboratory frame, which means that the nucleons are assumed to be at rest. For $\sigma_{\mathrm{tot}}^{(K^+N)}$ we take the parametrizations implemented in the Giessen transport model (GiBUU)~\cite{Buss:2011mx,LarionovPC}.

In the plane-wave limit, where the $KN$ interaction is neglected, $\phi^*_{>}(\vec{p}_K,\vec{r}) \raw \exp{(-i \vec{p}_K \cdot \vec{r})}$. In this limit, as we do not consider in-medium modifications of the $\Gamma_{i; NK^+}^\mu$, the nuclear current becomes
\begin{equation}
{\cal A}^\mu_{K^+}(q,p_K) \raw 
F_p(|\vec{q}-\vec{p}_K|) {\cal J}^\mu_{pK^+}(q,p_K) + 
F_n(|\vec{q}-\vec{p}_K|) {\cal J}^\mu_{nK^+}(q,p_K) 
\label{eq:APW}
\end{equation}
where $F_p$($F_n$) is the proton(neutron) nuclear form factor given by the Fourier transform of the corresponding density.

\section{Formalism for $K^-$ coherent production}
\label{sec:formanu}

\subsection{Single antikaon production model}
\label{subsec:microanu}

For the elementary process $\bar{\nu}_{l} \, p(n) \rightarrow l^+ \, K^- \, p(n)$ close to threshold, the relevant mechanisms can also be obtained  from chiral SU(3) Lagrangians~\cite{Alam:2012zz}. As for $\nu_{l} \, p(n) \rightarrow l^- \, K^+ \, p(n)$, contact term, kaon pole and $\pi$, $\eta$ in-flight contributions to the hadronic current are present but now the $\Lambda$ and $\Sigma$ hyperons appear in the s-channel. The structure of these amplitudes close to threshold is fully defined by chiral symmetry, with the couplings determined from semileptonic decays. As for $K^+$ production, the $q^2$ dependence is parametrized by a global dipole form factor $F(q^2)=(1-q^2/M_F^2)^{-2}$, with $M_F=1$~GeV, which we will keep fixed in this study. In pion production reactions, the excitation of the spin-3/2 $\Delta(1232)$ plays a dominant role at relatively low excitation energies ($\sim 200$~MeV). Therefore, the corresponding state of the baryon decuplet $\Sigma^*(1385)$ that couples to $N\bar{K}$ should be considered here. The vector and axial $N-\Sigma^*$ form factors, which are not known, are related to the better known $N-\Delta(1232)$ ones using SU(3) rotations. As can be seen in Figs.~3 and 4 of Ref.~\cite{Alam:2012zz}, the largest contribution to the cross section comes from the contact term. The small contribution from the $\Sigma^*$, contrasting with the dominance of $\Delta$ in the pion case, can be explained by the fact that the $\Sigma^*$ is below the kaon production threshold~\cite{Alam:2012zz}. 

\begin{figure}[h!]
\begin{center}
\includegraphics[width=0.9\textwidth]{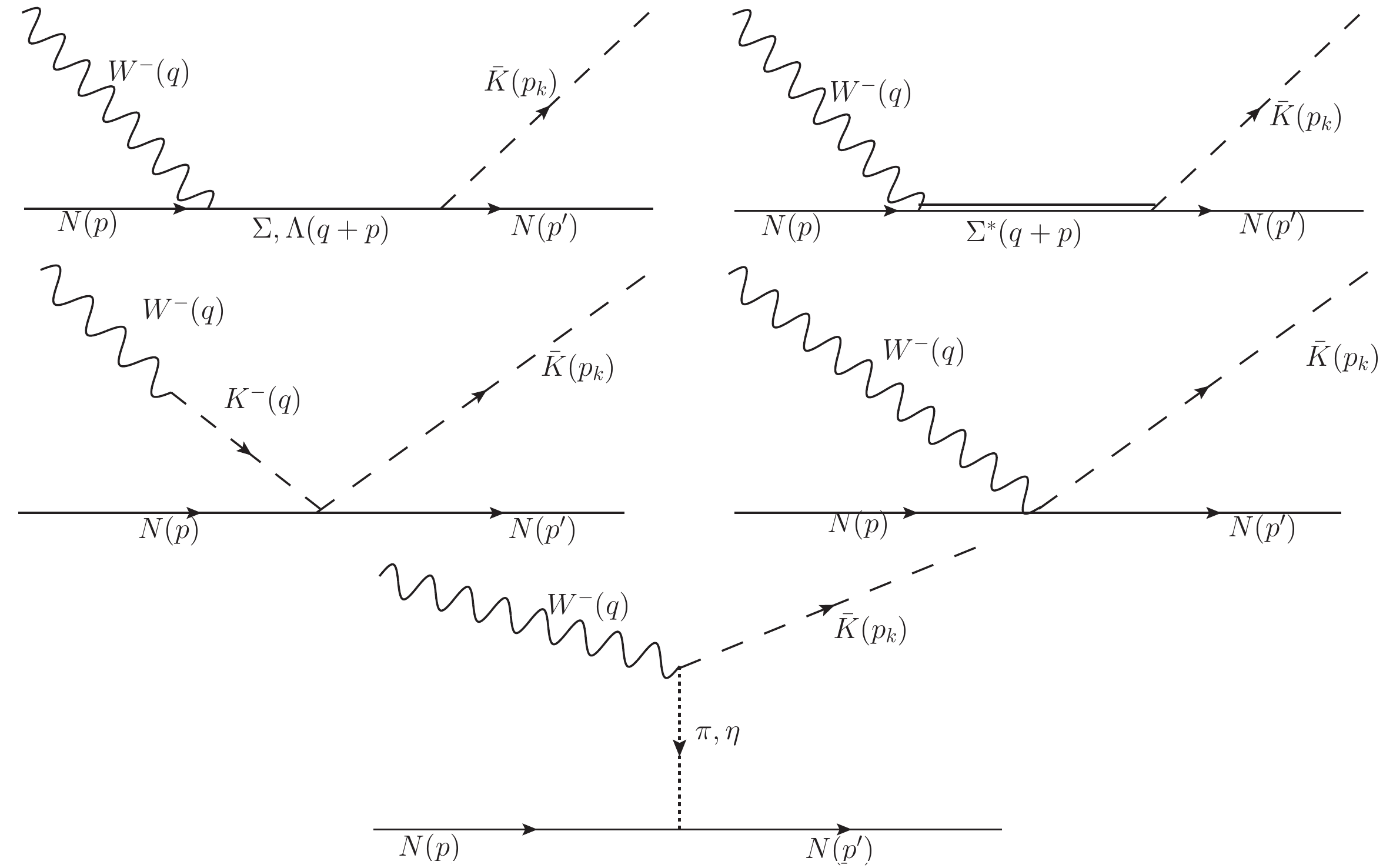}
\caption{\label{fig:diagsanu}
  Feynman diagrams for $W^- N \to  N \bar{K}$. First row:  s-channel $\Sigma, \Lambda$ and $\Sigma^*$ exchange terms; second row: contact (CT) and kaon pole (KP) terms; last row: $\pi$ and $\eta$ in flight ($\pi$P, $\eta$P) terms.} 
\end{center}
\end{figure}

\subsection{The coherent reaction}
\label{subsec:coheanu}

The formalism outlined in \ref{subsec:cohenu} for reaction (\ref{eq:reac1}) remains valid for (\ref{eq:reac2}) with a few modifications. Obviously, $K^-$ instead of $K^+$ should be understood in Eqs.~(\ref{eq:sec},\ref{eq:A},\ref{eq:J}). Now the index $i$ refers to all the possible mechanisms in Fig.~\ref{fig:diagsanu}; $\Gamma_{i; NK^-}^\mu$ can be obtained from the expressions in the Appendix of Ref.~\cite{Alam:2012zz}. As we have antineutrinos instead of neutrinos, the sign of the imaginary part in the leptonic tensor [Eq.~(\ref{eq:lep})] should be changed. In this model, the $\Sigma^*(1385)$ propagation is treated locally. Indeed, the $\Sigma^*$ momentum is well defined via the prescription that assigns a fixed momentum to the initial and final nucleons. In Ref.~\cite{Leitner:2009ph} this constrain was relaxed for the $\Delta(1232)$ in weak coherent pion production.  It was found that nonlocalities in the $\Delta$ propagation cause a reduction of the cross section at low energies. A similar result was obtained by Nakamura et al.~\cite{Nakamura:2009iq} with a different formalism. Being the $\Sigma^*$ heavier than the $\Delta$ and by far not as relevant, we expect any consequence from its non-local propagation in nuclei to be numerically minor.

The $\bar{K}$ interaction in the nuclear medium differs considerably from the $K$ one because of the more involved $\bar{K}$ interaction, with several channels ($\bar{K}N$, $\pi Y$, $\eta Y$, $Y=\Lambda,\,\Sigma$) open at low energies. For the $\bar{K}$ optical potential, we take the one developed in Ref.~\cite{Ramos:1999ku} based on a chiral unitary model in coupled channels for the  s-wave $\bar{K}N$ interaction~\cite{Oset:1997it} including medium effects such as Fermi motion, Pauli blocking and dressing of meson propagators with particle-hole and $\Delta$-hole excitations. A p-wave contribution from the excitation of $Y$-hole pairs [$Y=\Lambda,\,\Sigma,\,\Sigma^*(1385)$] is also included~\footnote{When solving the Klein-Gordon equation we treat this p-wave part as local.}. At $\rho = \rho_0$ this $V_{\mathrm{opt}}$ is attractive at low kaon momenta, becoming repulsive at $\sim 500$~MeV$/c$. The range of applicability of $V_{\mathrm{opt}}$ restricts our calculation to $|\vec{p}_{\bar{K}}| \leq 1$~GeV$/c$.

\section{Results}
\label{sec:res}

\subsection{$\nu_{\mu} \, {}^{A}\! Z_{\text{gs}} \to \mu^- \, {}^{A}\! Z_\text{gs} \, K^+$}
\label{subsec:nures}

\begin{figure}[h!]
\begin{center}
  \makebox[0pt]{\includegraphics[width=0.5\textwidth]{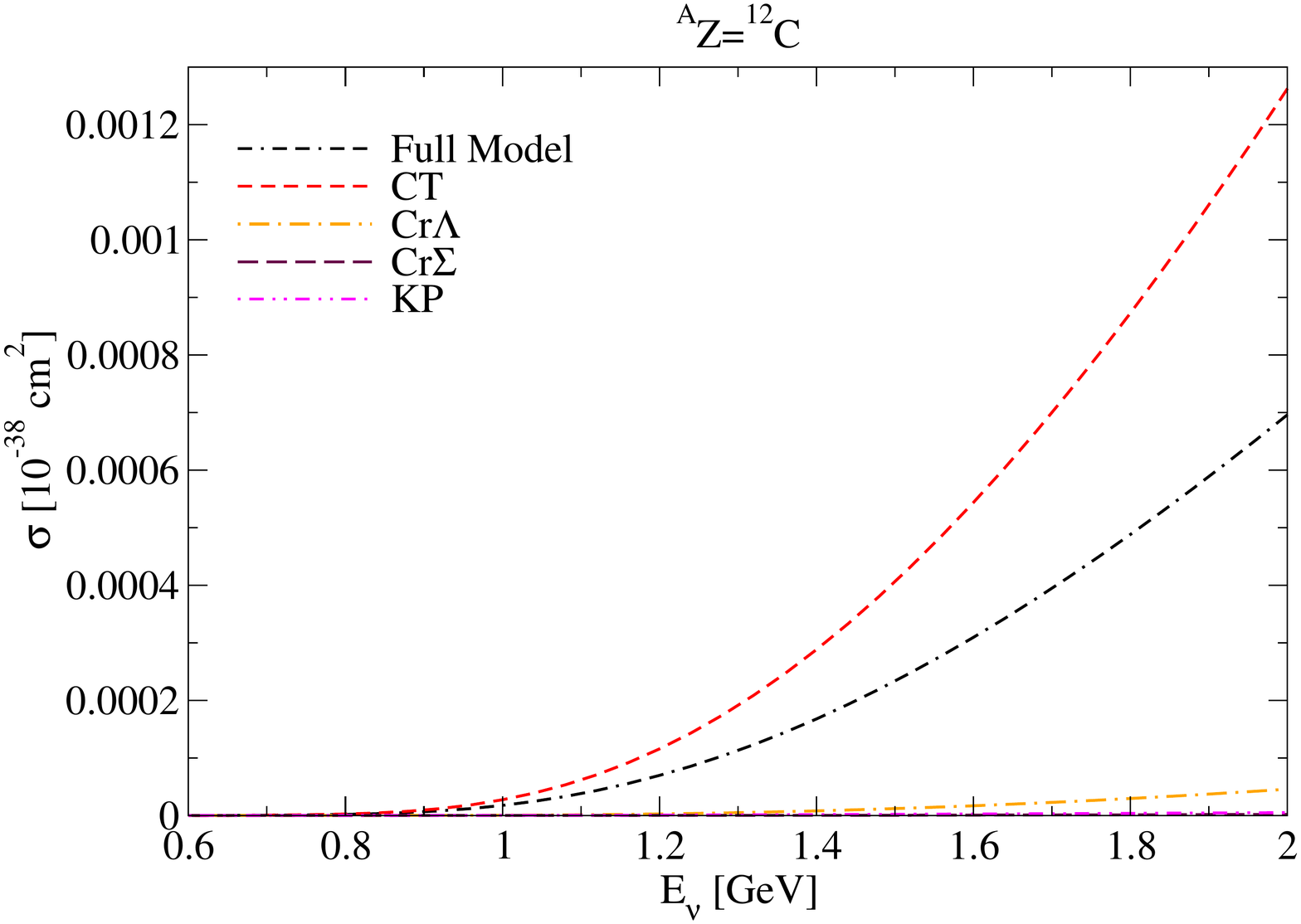}\hspace{0.cm}
              \includegraphics[width=0.5\textwidth]{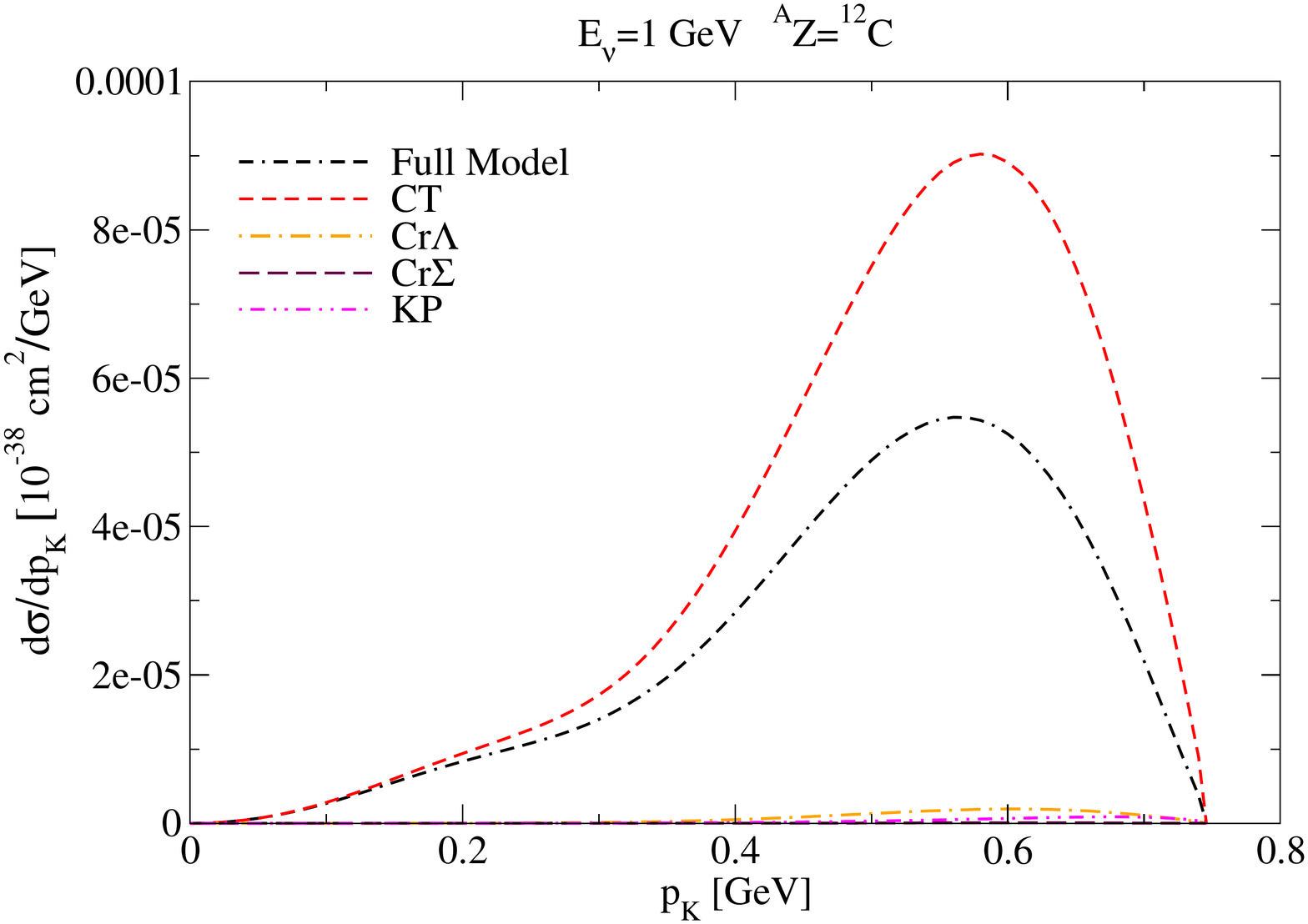}}
\end{center}
\caption{\label{fig:mechs} (Color online) Contribution of different kaon production mechanisms to the coherent reaction on $^{12}$C. Left panel: total cross section as a function of the neutrino energy. Right panel: kaon momentum distribution for 1~GeV neutrinos. Kaon distortion is not taken into account.}
\end{figure}

In Fig.~\ref{fig:mechs}, for $^{12}$C, we show the contribution of the different mechanisms to the integrated cross section and to the kaon momentum distribution at $E_\nu= 1$~GeV, ignoring kaon distortion. The cross section is evaluated in the validity range of the kaon production model on the nucleon accepted in Ref.~\cite{RafiAlam:2010kf}. The largest contribution arises from the CT. The rest of the mechanisms, mainly Cr$\Lambda$, account for less than 1/25 of the CT at $E_\nu =2$~GeV. Nevertheless, there is a strong destructive interference that reduces the cross section considerably. This pattern, already present in the elementary reaction (see Figs.~2,4 of Ref.~\cite{RafiAlam:2010kf}), is enhanced by the kinematics of coherent scattering that favors low momentum transfers. With our approximation for the nucleon momenta discussed in \ref{subsec:cohenu}, the contribution from $\pi$P and $\eta$P vanishes exactly.     

The reaction cross section turns out to be quite small. 
At $E_{\nu}=2$~GeV, the cross section per nucleon is $\sim 40$ times smaller than the one on free nucleons averaged over protons and neutrons (compare Fig.~\ref{fig:mechs} to Figs.~2,4 of Ref.~\cite{RafiAlam:2010kf}). This is the consequence of producing a rather heavy particle like a kaon at low energies, leaving the final nucleus in its ground state. 
Indeed, the momentum transferred to the nucleus should be as small as possible, otherwise the nuclear form factors, that appear squared in the cross sections [see Eqs.~(\ref{eq:APW},\ref{eq:sec})], are drastically reduced. In our case 
$|\vec{q}-\vec{p}_K| \ge q_0 - | \vec{p}_K| \approx \sqrt{m_K^2 + \vec{p}_K^{\, 2}} - | \vec{p}_K|$, which is large at moderate kaon momenta. In particular, at $|\vec{p}_K|=0$ it is equal to $m_K$, decreasing for larger $|\vec{p}_K|$, which are favored as can be seen in the right panel of Fig.~\ref{fig:mechs}. To illustrate the impact of the kaon mass we have reduced it by a factor 2, finding that the integrated cross section is increased by a factor 10 at $E_\nu=1$~GeV and 68 at $E_\nu=1.5$~GeV. Another consequence of the large momentum transfers that are typical for this reaction at low energies is the large sensitivity to the nuclear density distributions. 

\begin{figure}[h!]
\begin{center}
\makebox[0pt]{\includegraphics[width=0.5\textwidth]{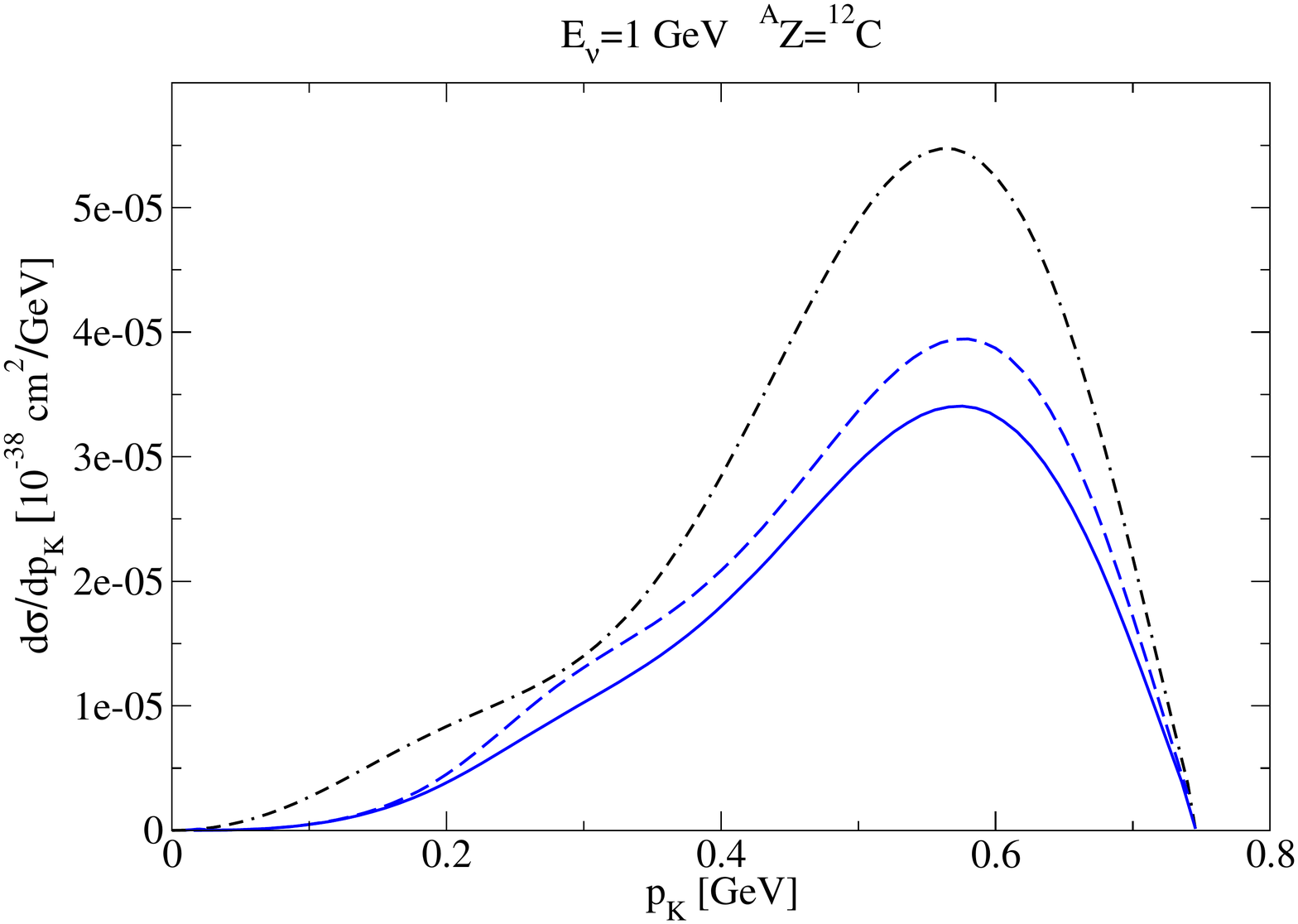}\hspace{0.cm}
              \includegraphics[width=0.5\textwidth]{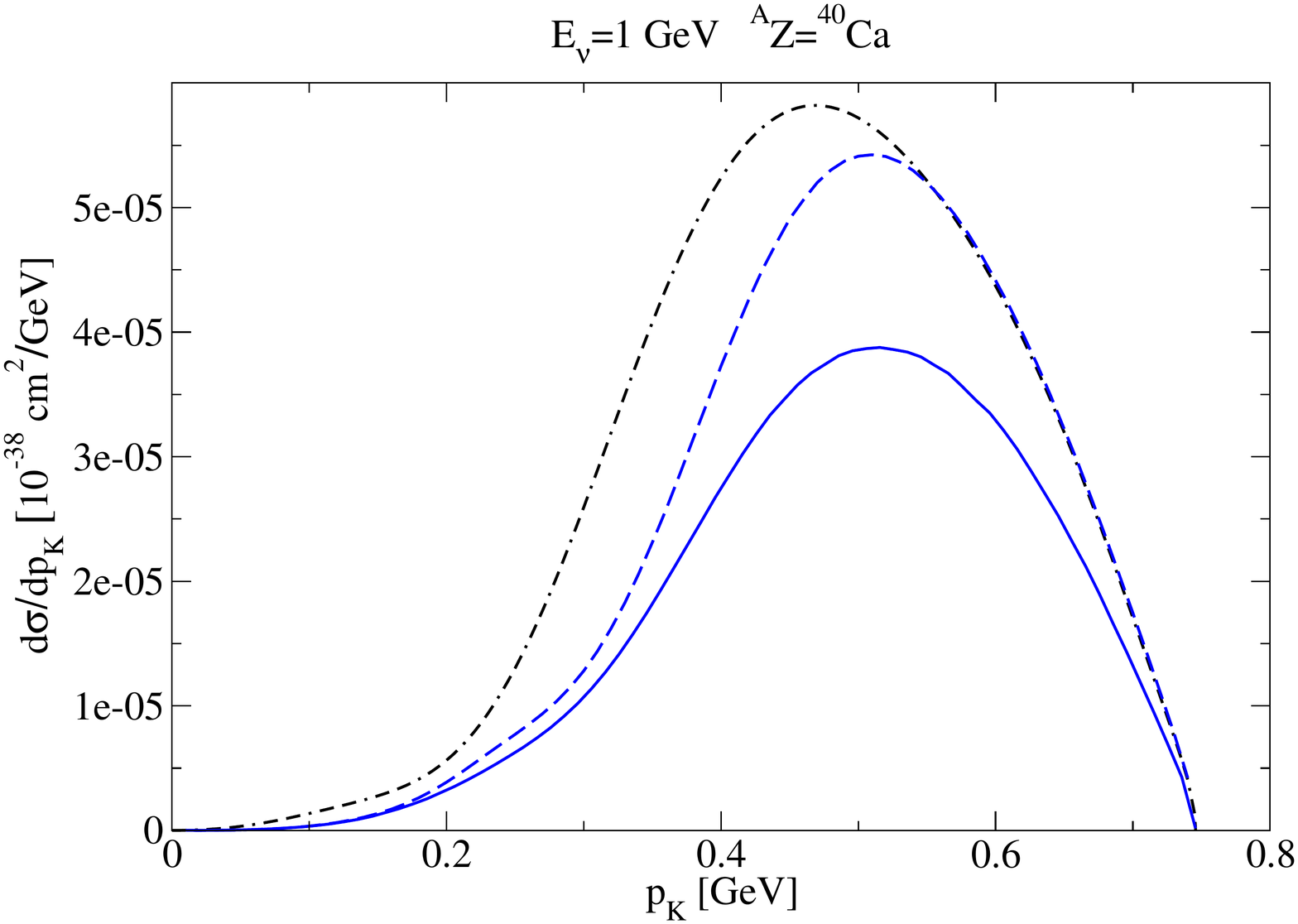}}

\end{center}
\caption{\label{fig:sige}
(Color online). 
  Differential cross section as a function of the outgoing kaon
  momentum at $E_\nu=1$~GeV for two different nuclei. The dash-dotted line 
is obtained with the full model for kaon plane waves. The other two incorporate 
kaon distortion with only the real part of $V_{\mathrm{opt}}$ (dashed line) and including also the absorptive term of Eq.~(\ref{eq:opt}) (solid line).}
\end{figure}

The impact of the distortion of the kaon wave function on the kaon momentum distributions is shown in Fig.~\ref{fig:sige} at $E_\nu = 1$~GeV and for two different nuclei ($^{12}$C, $^{40}$Ca). In presence of the optical potential there is a reduction of the cross section even when only the real part is taken into account. The imaginary part of the potential causes a further reduction which is larger for the heavier nucleus as one would expect.

\begin{figure}[h!]
\begin{center}
\makebox[0pt]{\includegraphics[width=0.5\textwidth]{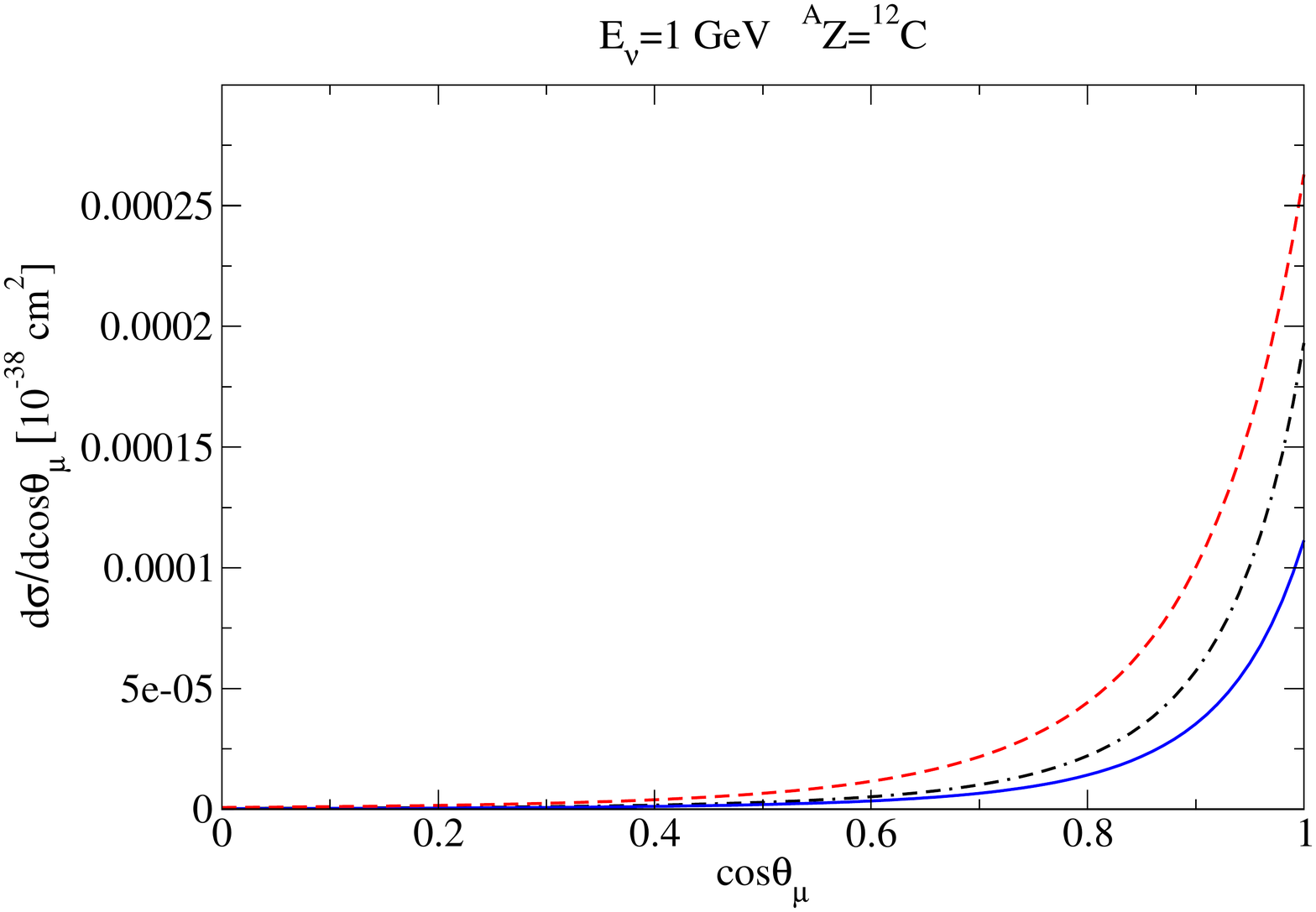}\hspace{0.cm}
              \includegraphics[width=0.5\textwidth]{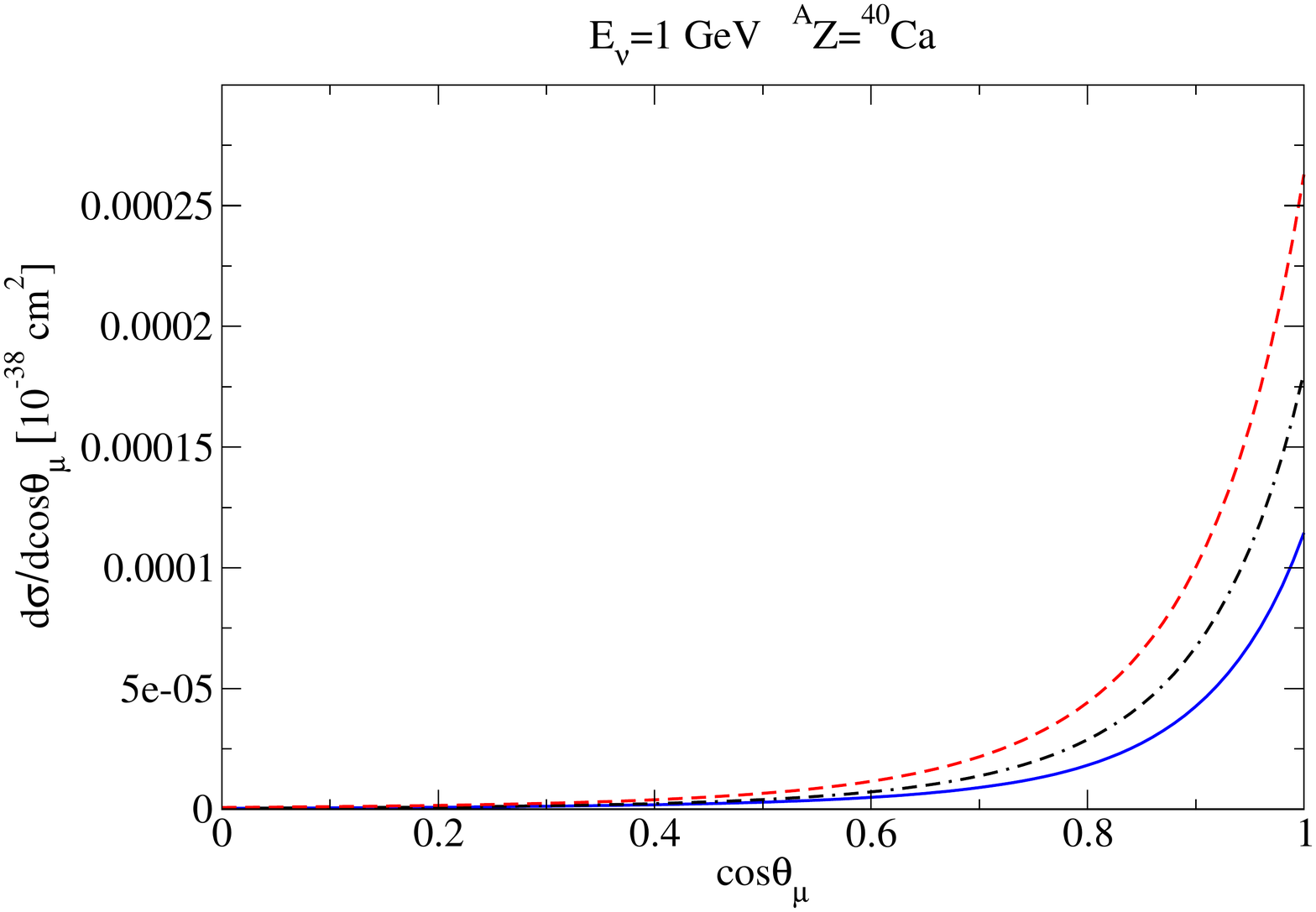}}

\end{center}
\caption{\label{fig:dkmu} (Color online) Muon angular distribution in the Laboratory system at $E_\nu=1$~GeV for two different nuclei. Results are shown for the largest CT mechanism without kaon distortion (dashed lines) and for the full model without (dash-dotted lines) and with kaon distortion (solid lines).}
\end{figure}

We now turn our attention to the outgoing lepton angular distribution shown in 
Fig.~\ref{fig:dkmu}. The reaction is very forward peaked. Furthermore the 
distribution profile is practically not affected by kaon distortion. 
Similar features have already been described in weak coherent pion production.

\begin{figure}[h!]
\begin{center}
\makebox[0pt]{\includegraphics[width=0.5\textwidth]{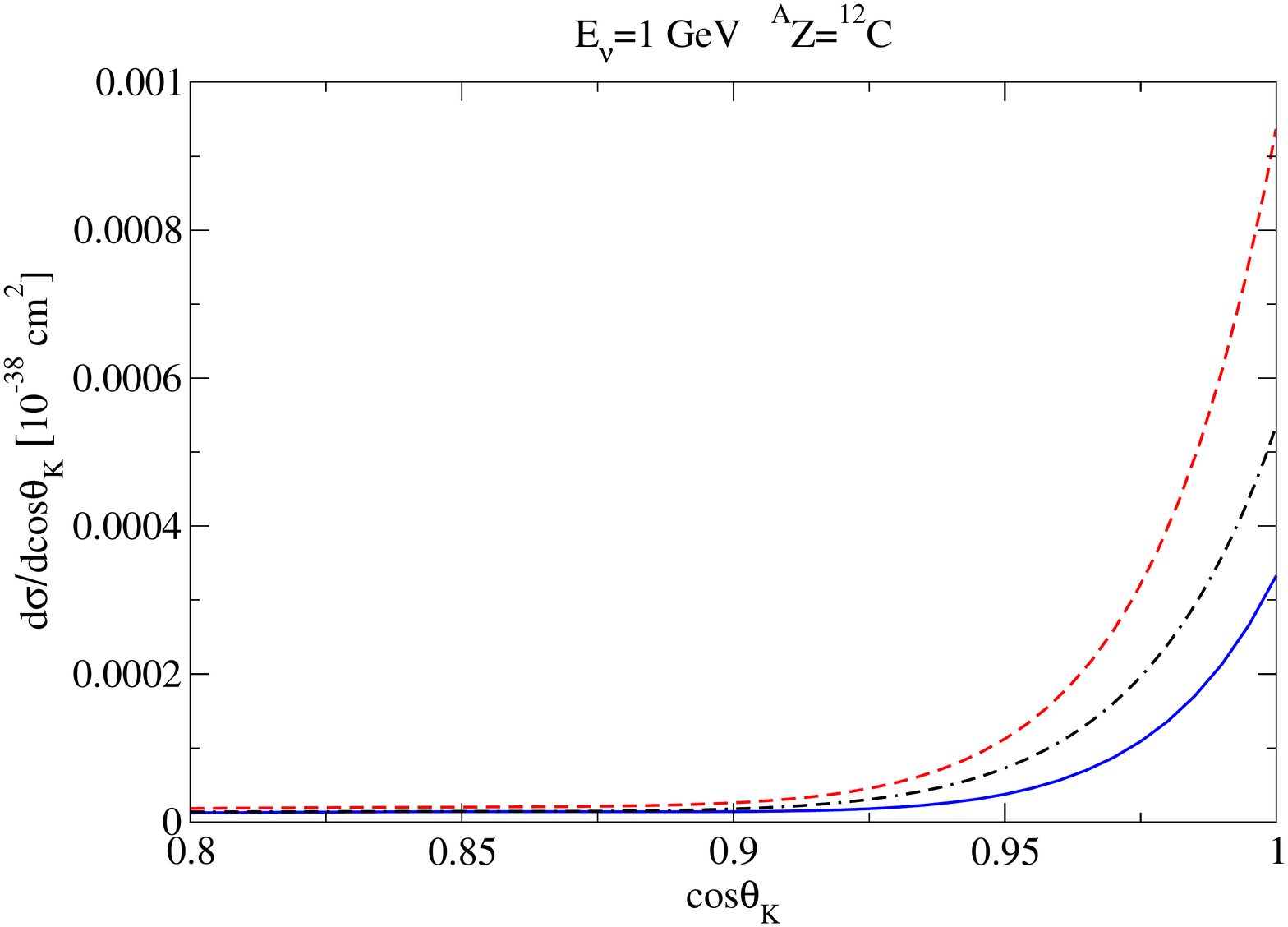}\hspace{0.cm}
              \includegraphics[width=0.5\textwidth]{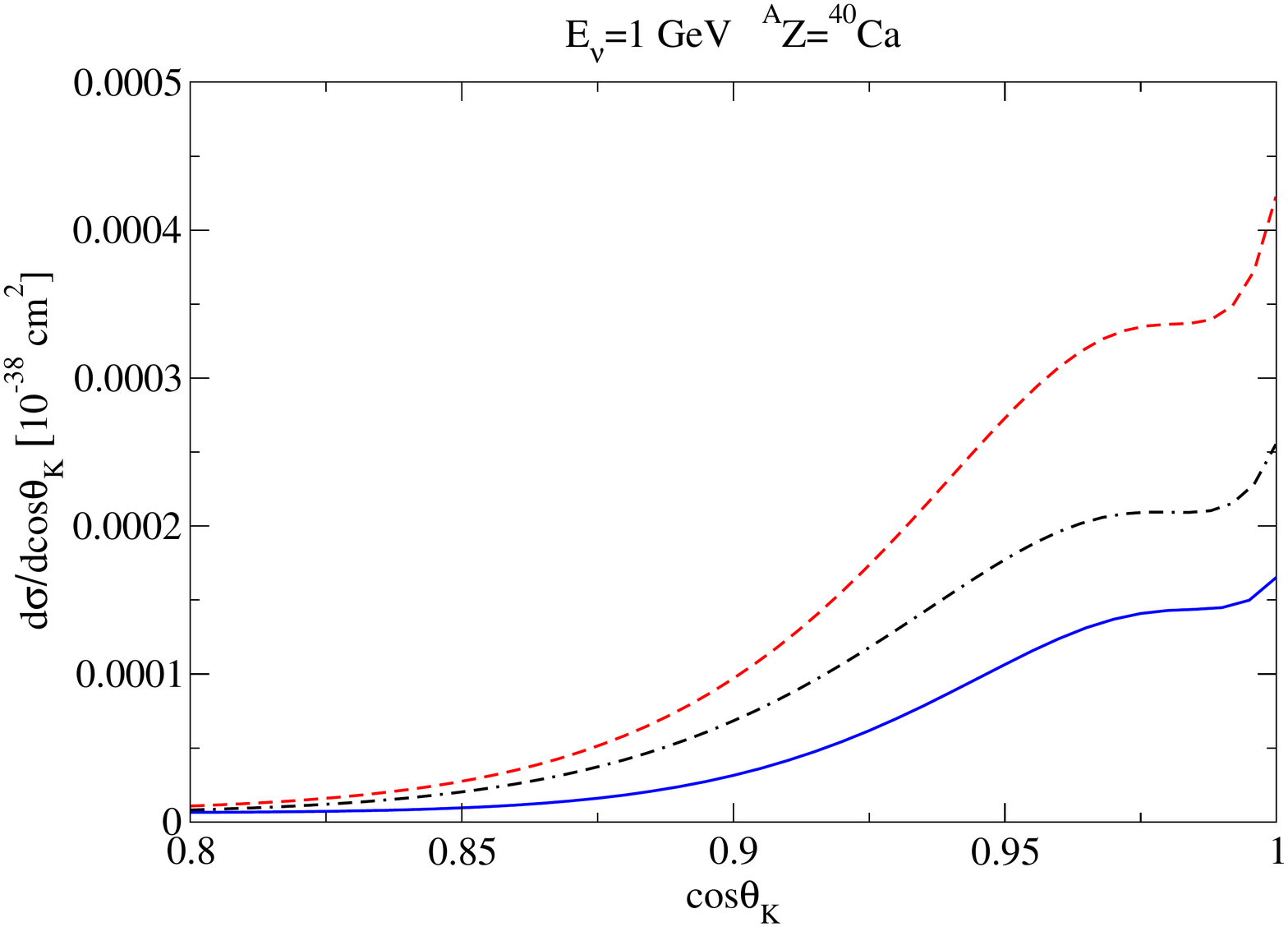}}
\end{center}
\caption{\label{fig:dkk} (Color online) Kaon angular distribution in the Laboratory system at $E_\nu=1$~GeV for two different nuclei. Lines have the same meaning as in Fig.~\ref{fig:dkmu}.}
\end{figure}

The outgoing kaon angular distributions shown in Fig.~\ref{fig:dkk} are also forward peaked, but considerably less in the case of $^{40}$Ca. At the first sight, this is in contradiction with the fact that heavier nuclei have narrower form factors. However, it is precisely because the narrow form factor of $^{40}$Ca that this distribution is sensitive to the second diffractive maximum and becomes wider. This diffractive pattern is smoothed by kaon distortion.

\begin{figure}[h!]
\begin{center}
\makebox[0pt]{\includegraphics[width=0.5\textwidth]{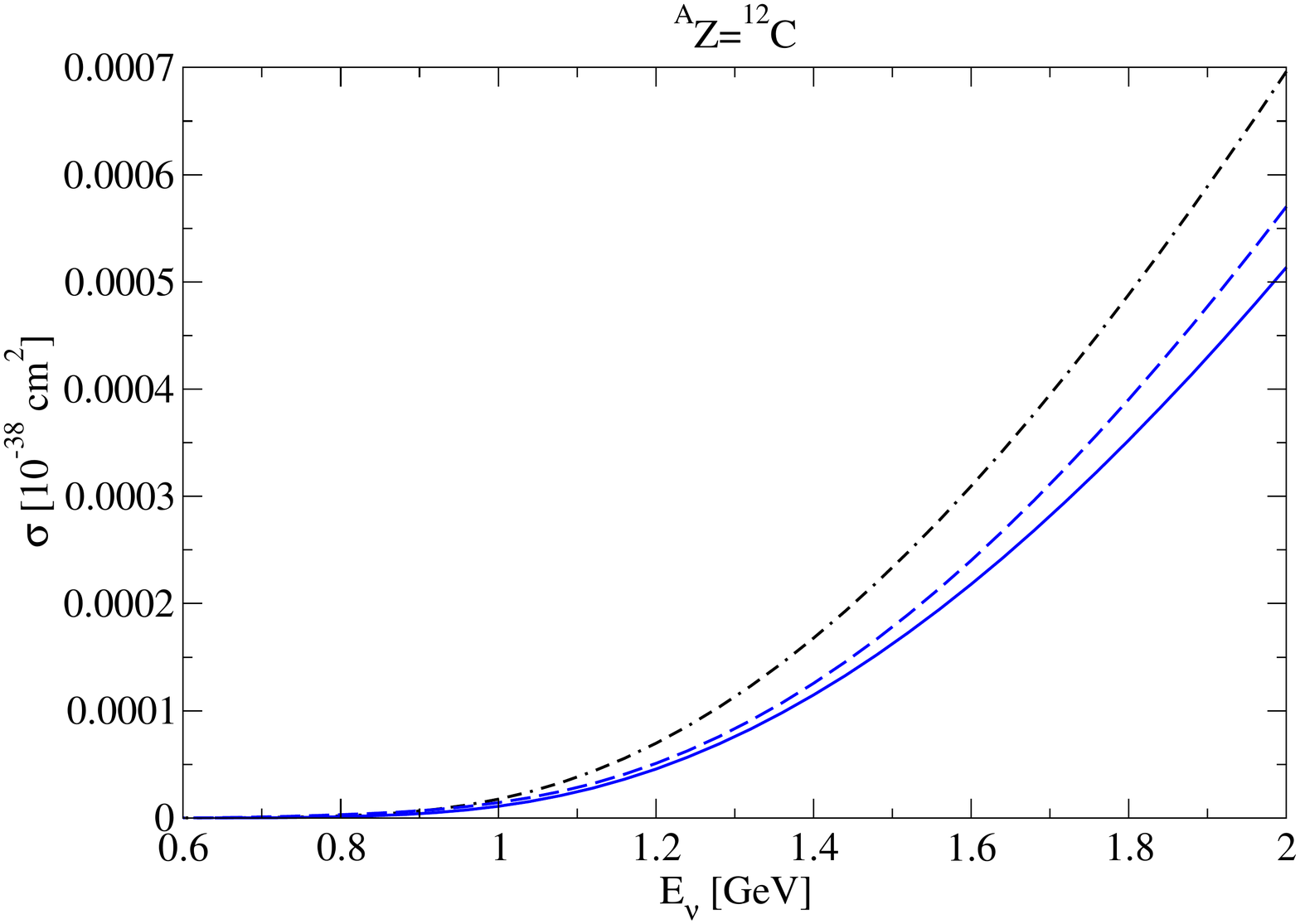}\hspace{0.cm}
              \includegraphics[width=0.5\textwidth]{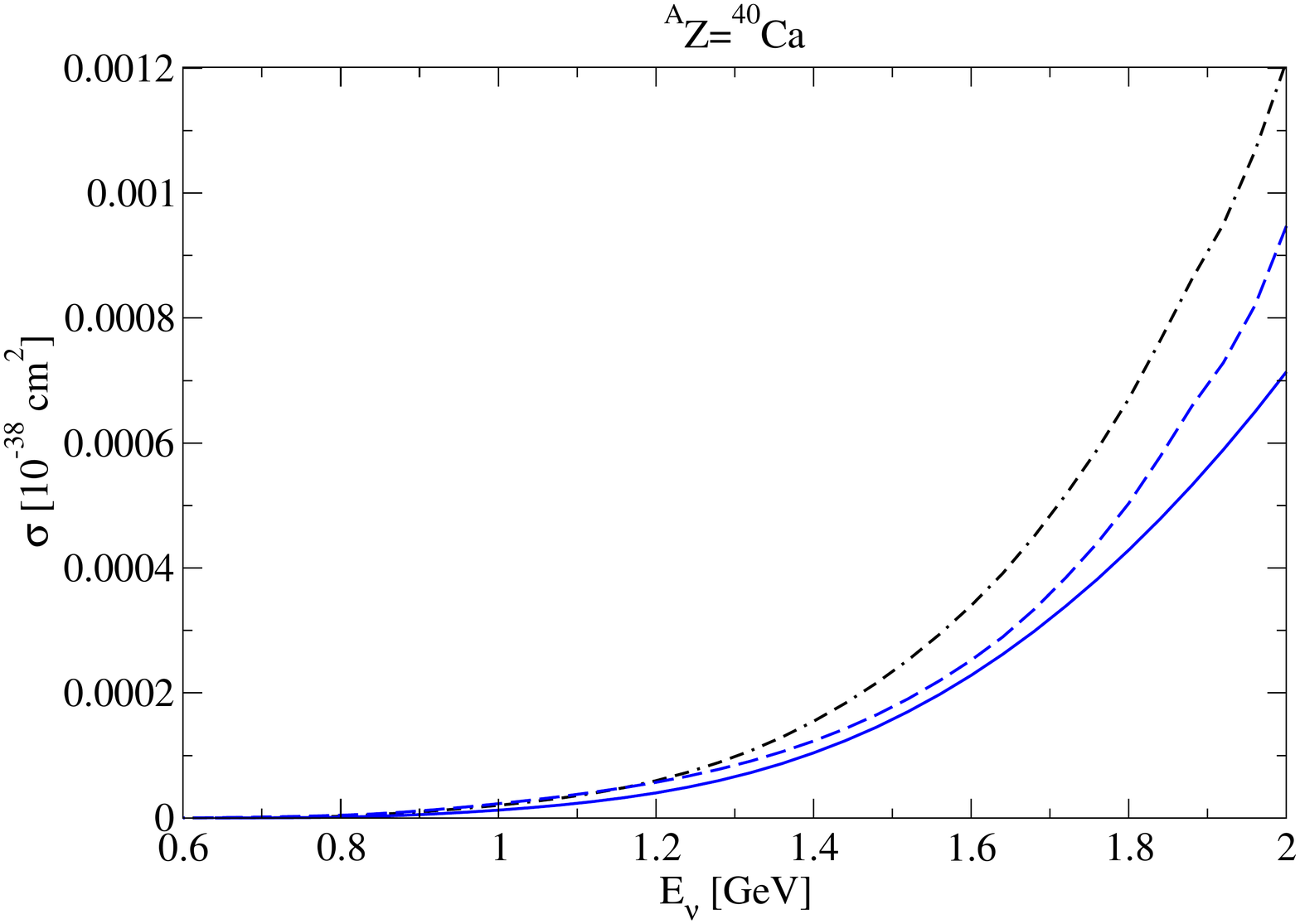}}
\end{center}
\caption{\label{fig:siged} (Color online) Integrated cross section as a function of the neutrino energy.  Dash-dotted line are obtained with the full model for kaon plane waves. The other two line styles denote results that incorporate kaon distortion with only the real part of $V_{\mathrm{opt}}$ (dashed line), and including also the absorptive term of Eq.~(\ref{eq:opt}) (solid line).}
\end{figure}

Next we discuss the energy dependence of the total cross section for $^{12}$C and $^{40}$Ca targets, given in Fig.~\ref{fig:siged}. As in Fig.~\ref{fig:sige} we show the effect of both the real and imaginary part of the kaon optical potential on the results. It is clear that the reduction caused by the absorptive term is not large but increases with energy and atomic number. 

Finally, we have investigated how the total cross section changes with the atomic and mass numbers of the target nuclei. The global factor in front of the dominant CT implies a dependence of the amplitude on the nucleon density $\sim \rho_n + 2\rho_p$, which suggests a quadratic dependence of the cross section on the variable $A + Z$. In practice, although an overall increase of $\sigma$ with $A+Z$ is observed, it is much slower than $(A + Z)^2$, even when the kaon distortion is neglected. Moreover, the actual trend for medium-size nuclei is quite irregular and likely due to the nontrivial interplay between the cross section increase when more nucleons are added to the system, the fact that heavier nuclei have narrower form factors, which causes a larger suppression of high $\vec{q} - \vec{p}_K$, and the contribution of secondary diffractive maxima.
\begin{figure}[h!]
  \begin{center}
    \includegraphics[scale=0.36]{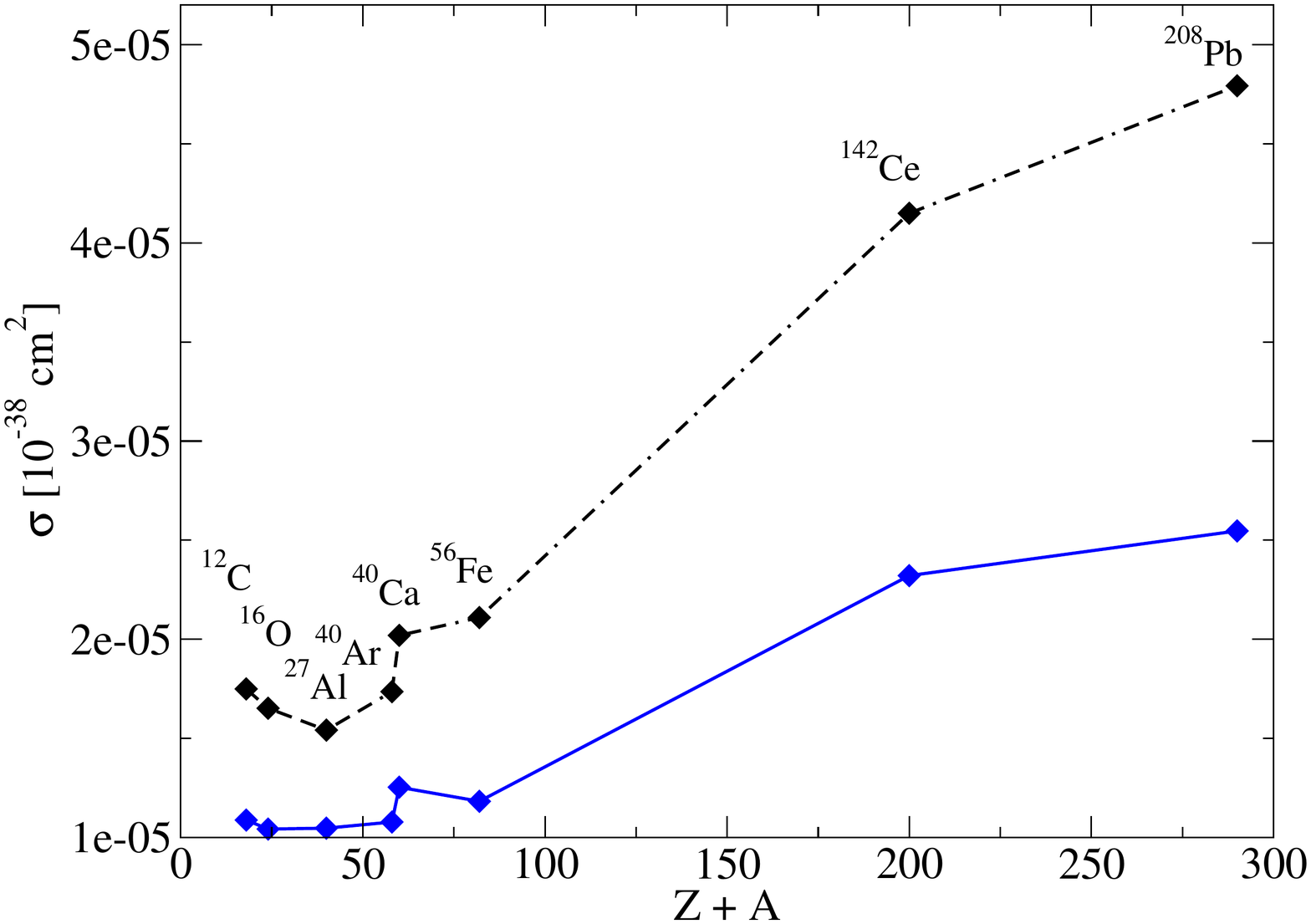}
  \end{center}
  \caption{\label{fig:zrange}  (Color
    online) Total cross section for $\nu_{\mu} \, {}^{A}\! Z_{\text{gs}} \to \mu^- \, {}^{A}\! Z_\text{gs} \, K^+$ 
as a function of $A+Z$ at $E_\nu =1$~GeV for several nuclei. The dashed (solid) line stands for the calculation without (with) kaon distortion.}
\end{figure}

\subsection{$\bar{\nu}_l  \, {}^{A}\! Z_{\text{gs}} \to l^+ \, {}^{A}\! Z_\text{gs} \, K^-$}
\label{subsec:anures}

First of all we present the contribution of the different reaction mechanisms to the integrated cross section (in the energy interval where the elementary model was considered to be valid in Ref.~\cite{Alam:2012zz}) and the kaon momentum distribution (for 1~GeV antineutrinos) on $^{12}$C. Antikaon distortion has been neglected. The interferences largely reduce the cross section from the otherwise dominant CT. The comparison with the cross sections on the nucleon given in  Ref.~\cite{Alam:2012zz} show a much stronger interference in the present (coherent) reaction. Another difference is that $\Sigma^*$ excitation is now the second largest piece, being as large as the full model around the maximum of the $|\vec{p}_K|$ distribution. With our choice of average momenta for the nucleons in the target, $\pi$P and $\eta$P currents are exactly zero.

As for $K^+$ coherent production, we find that at $E_{\bar{\nu}}=2$~GeV, the cross section per nucleon is $\sim$ 40 times smaller than the elementary one averaged over protons and neutrons. The explanation given in \ref{subsec:anures} in terms of the large kaon mass compared with the typical kaon momenta also applies here.  
\begin{figure}[h!]
\begin{center}
  \makebox[0pt]{\includegraphics[width=0.5\textwidth]{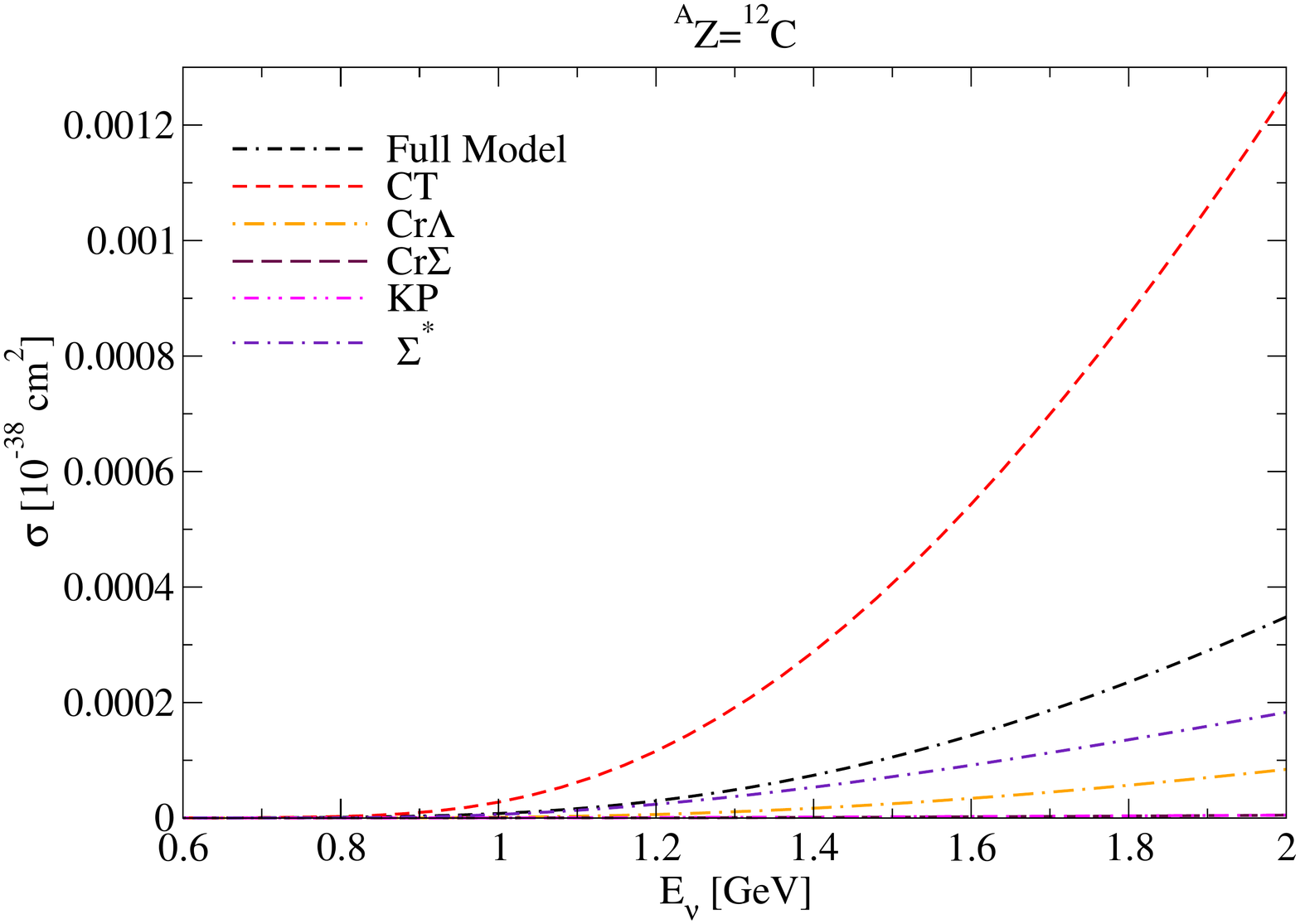}\hspace{0.cm}
              \includegraphics[width=0.5\textwidth]{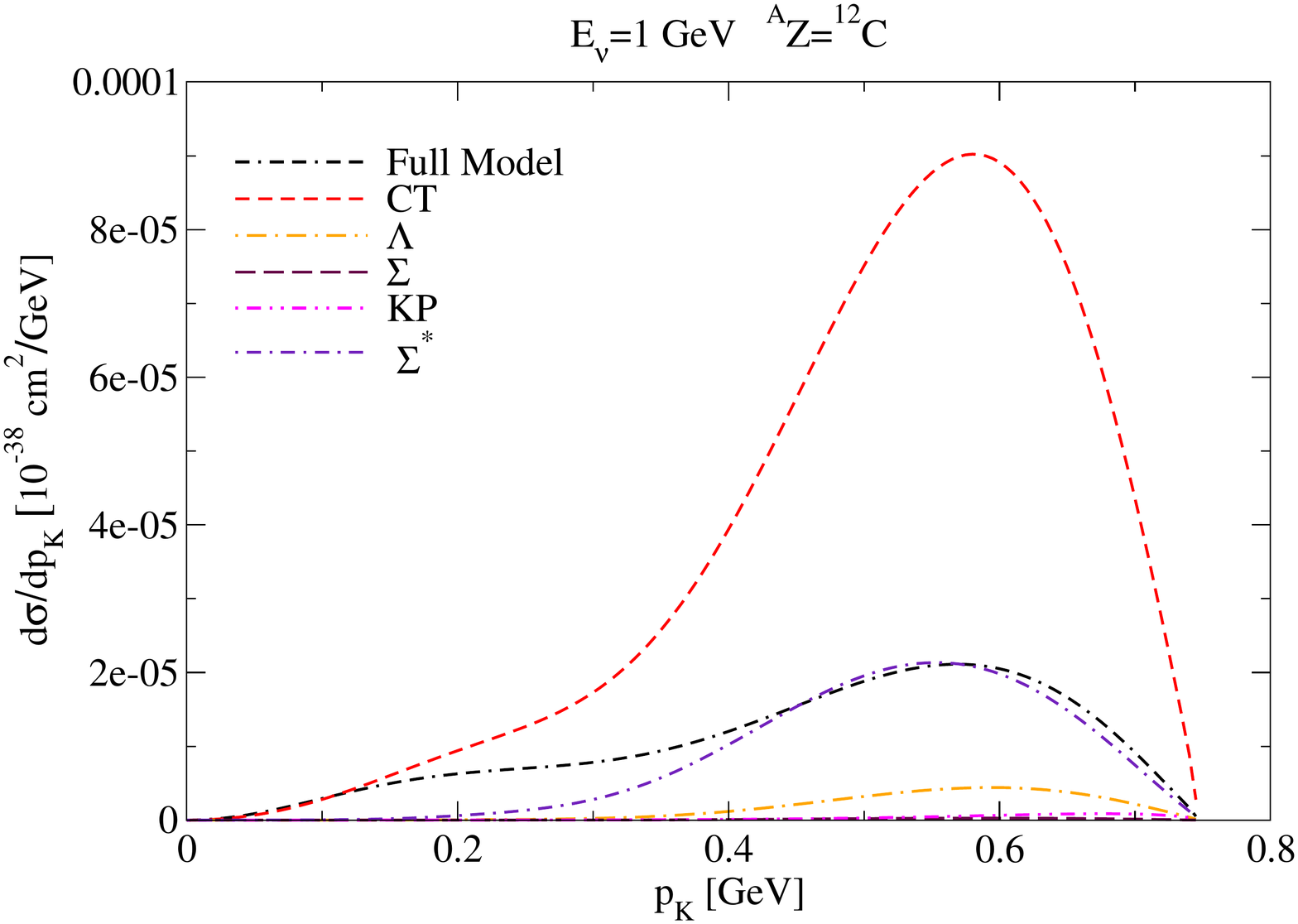}}
\end{center}
\caption{\label{fig:amechs} (Color online) Contribution of different $K^-$ production mechanisms to the coherent reaction on $^{12}$C. Left panel: total cross section as a function of the neutrino energy. Right panel: kaon momentum distribution for 1~GeV neutrinos. Antikaon distortion is not taken into account.}
\end{figure}

\begin{figure}[h!]
\begin{center}
\makebox[0pt]{\includegraphics[width=0.5\textwidth]{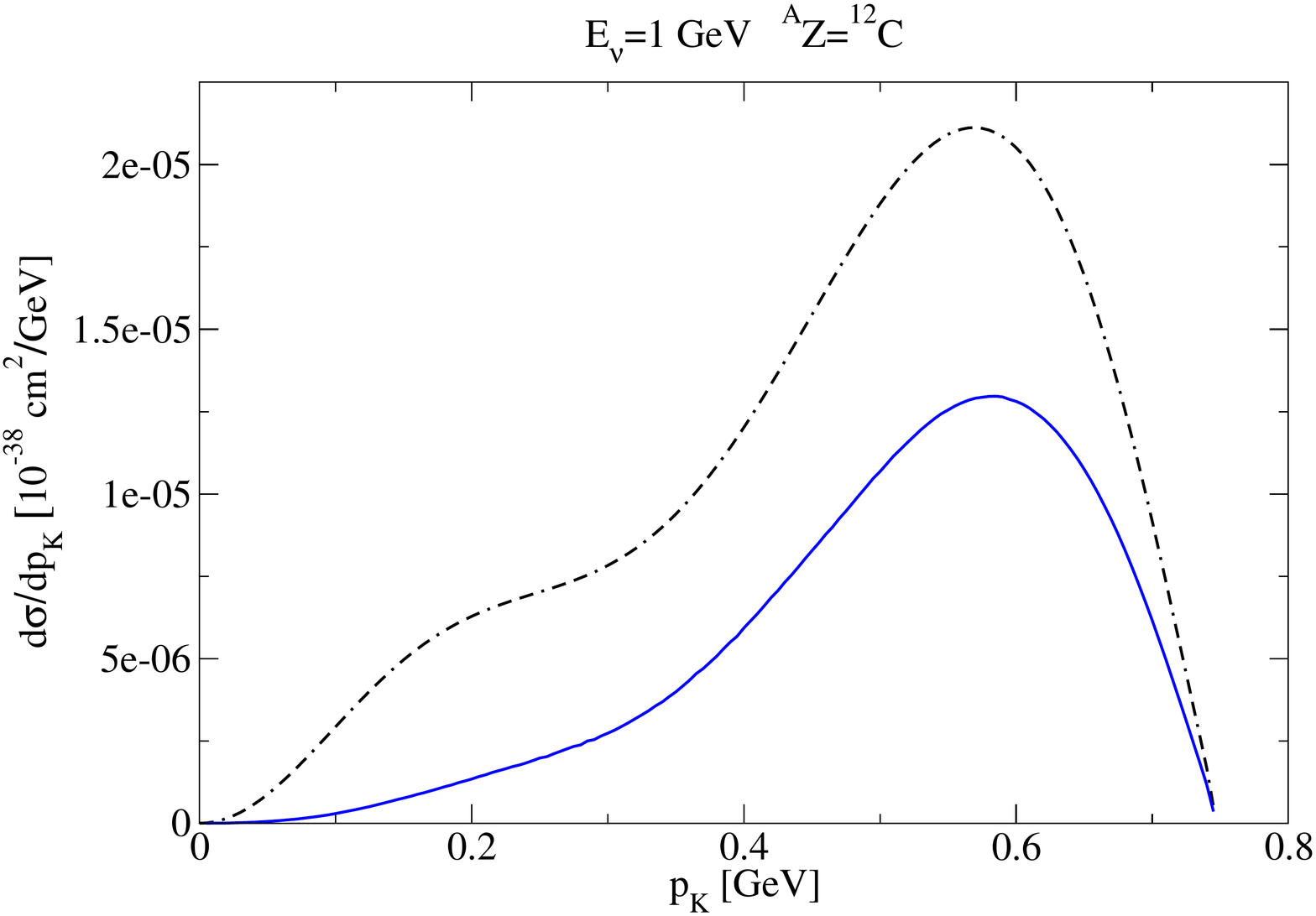}\hspace{0.cm}
              \includegraphics[width=0.5\textwidth]{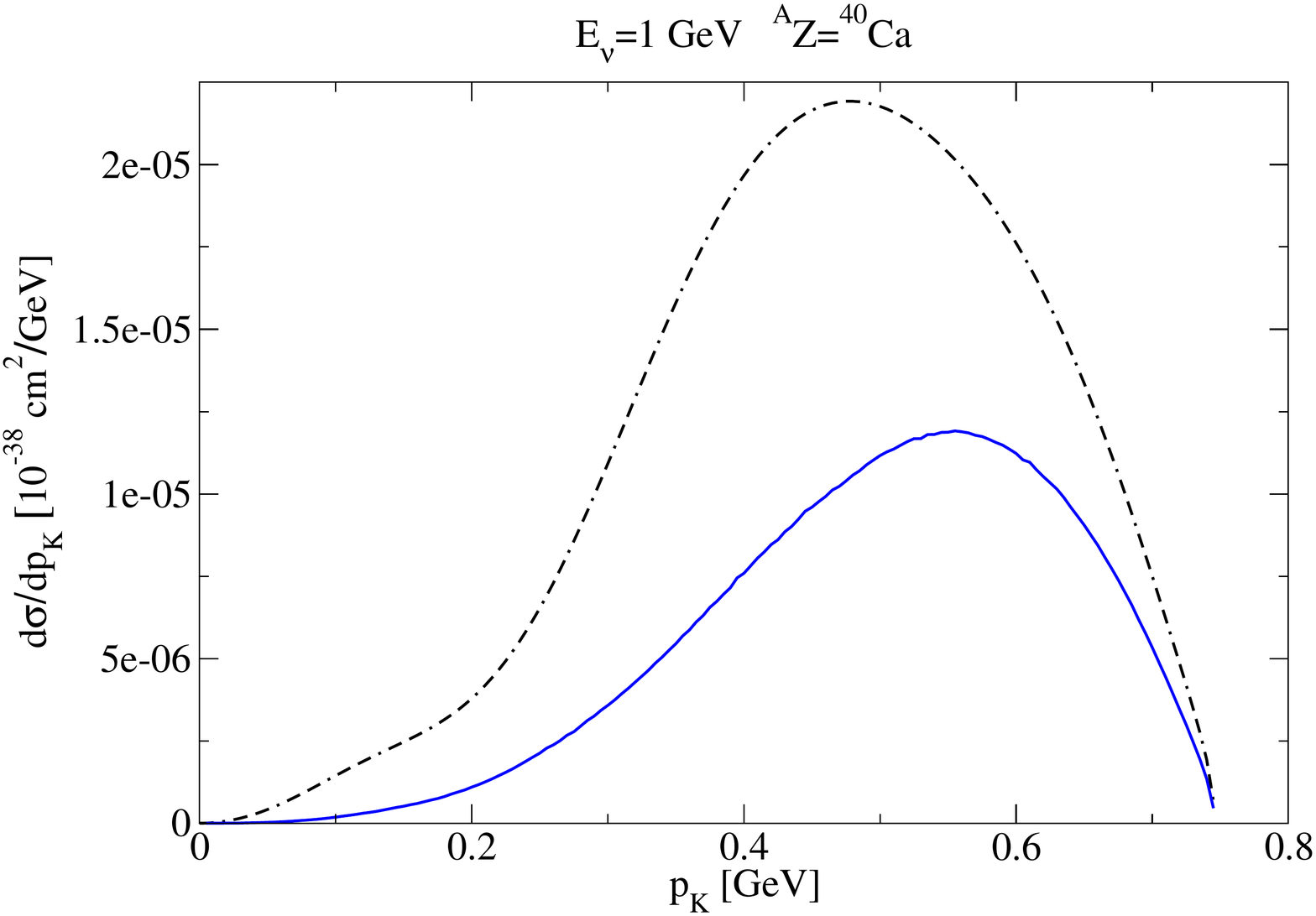}}

\end{center}
\caption{\label{fig:asige}
(Color online). 
  Differential cross section as a function of the outgoing antikaon
  momentum at $E_{\bar{\nu}}=1$~GeV for two different nuclei. The curves are obtained for the full model without (dash-dotted lines) and with antikaon distortion (solid lines).}
\end{figure}

The distortion of the outgoing $K^-$ waves makes kaon-momentum distribution smoother and reduces the cross section values (see Fig.~\ref{fig:asige}). This reduction is larger than for $K^+$ coherent production due to the stronger $\bar{K}$ interaction in the nuclear medium. 
\begin{figure}[h!]
\begin{center}
\makebox[0pt]{\includegraphics[width=0.5\textwidth]{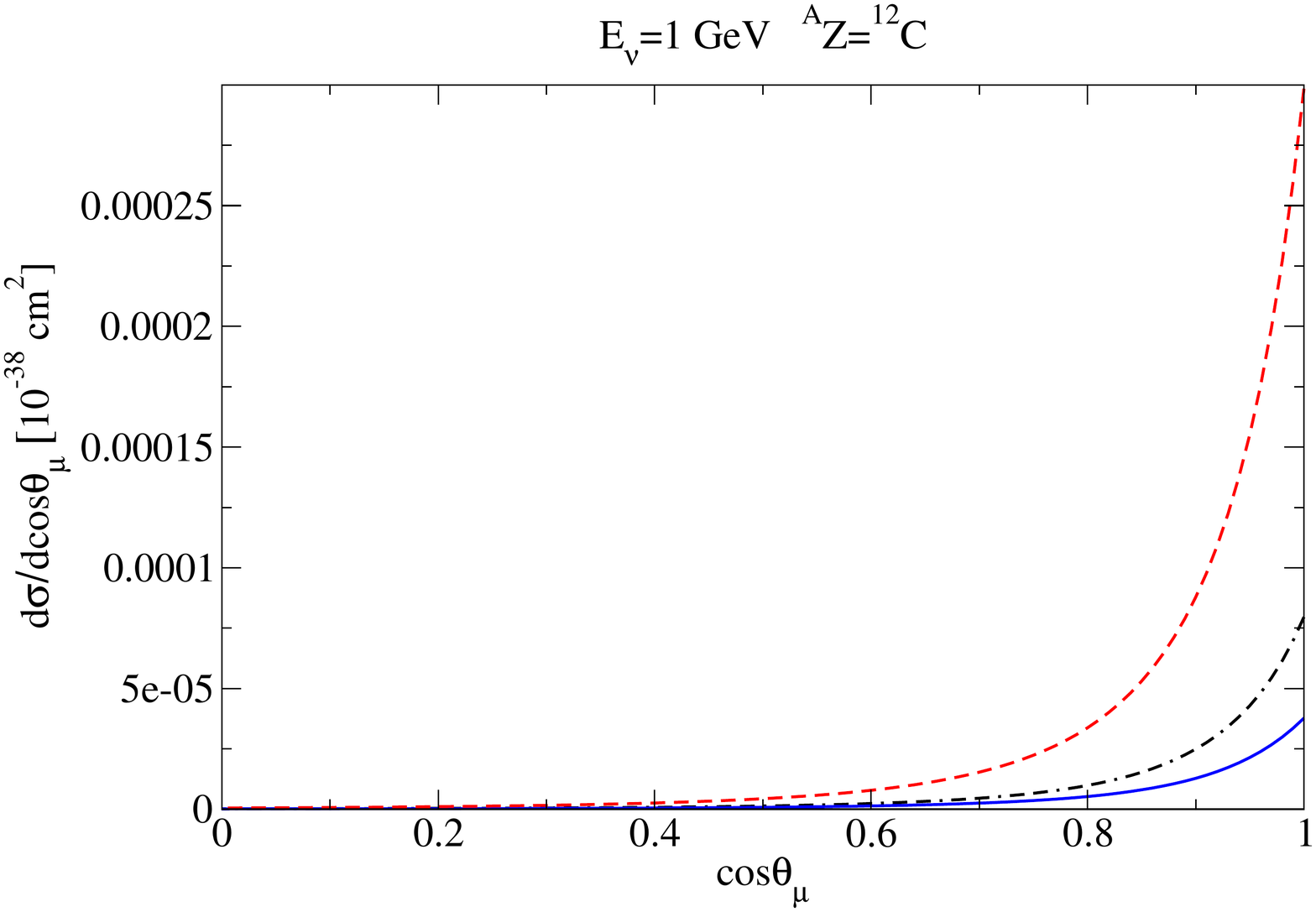}\hspace{0.cm}
              \includegraphics[width=0.5\textwidth]{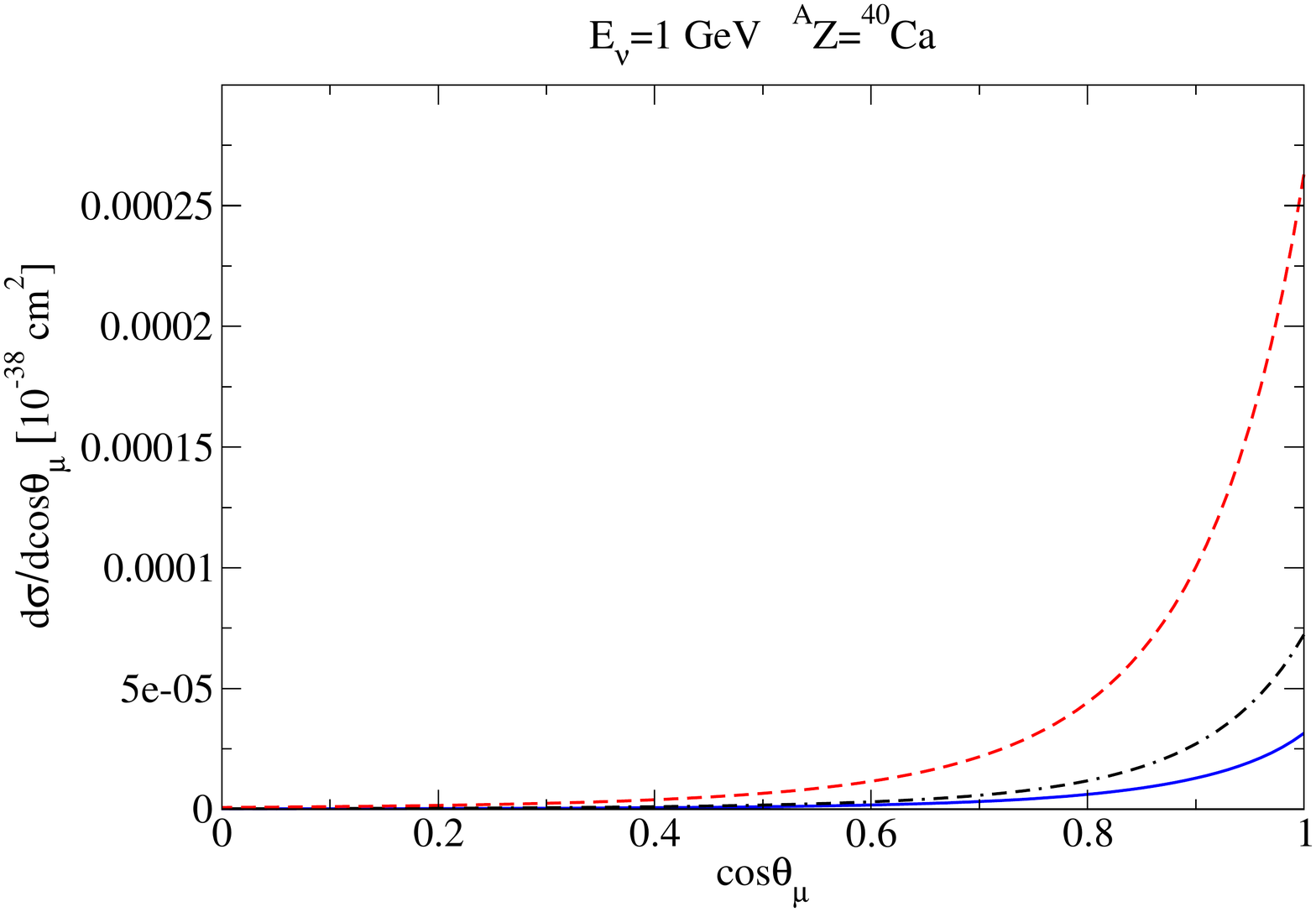}}

\end{center}
\caption{\label{fig:adkmu} (Color online) Muon angular distribution in the Laboratory frame at $E_{\bar{\nu}}=1$~GeV for two different nuclei. Results are shown for the largest CT mechanism without antikaon distortion (dashed lines) and for the full model without (dash-dotted lines) and with antikaon distortion (solid lines).}
\end{figure}
The forward-peaked angular distributions of outgoing leptons and kaons, characteristic for coherent scattering are again present, as can be observed in Figs.~\ref{fig:adkmu},\ref{fig:adkk}. They are very narrow for the CT mechanism alone, becoming wider for the full model and even more after the kaon distortion is turned on. The smoothening effect of the distortion is clearly seen in the kaon angular distribution for $^{40}$Ca [right panel of Fig.~\ref{fig:adkk}]. 
\begin{figure}[h!]
\begin{center}
\makebox[0pt]{\includegraphics[width=0.5\textwidth]{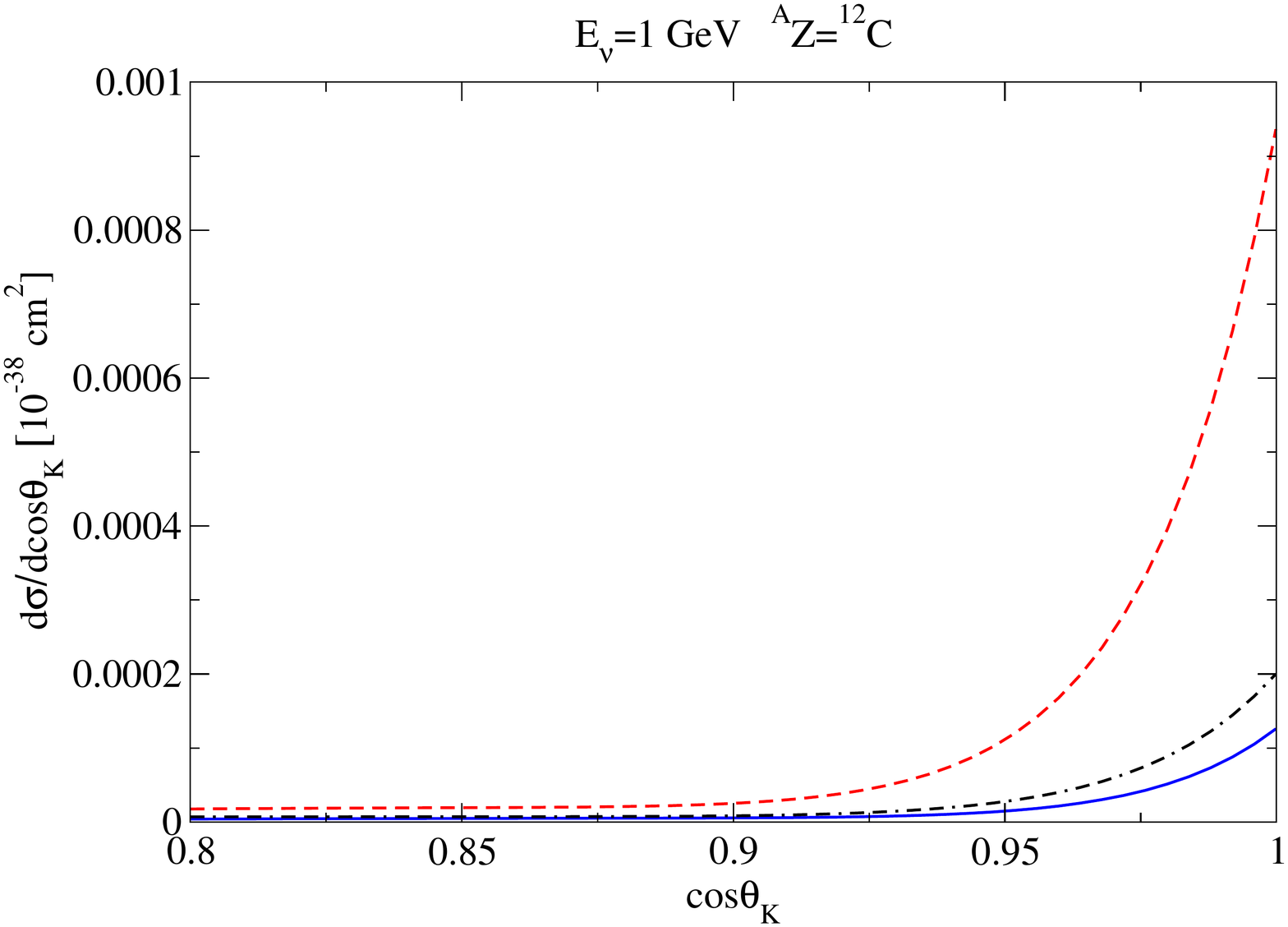}\hspace{0.cm}
              \includegraphics[width=0.5\textwidth]{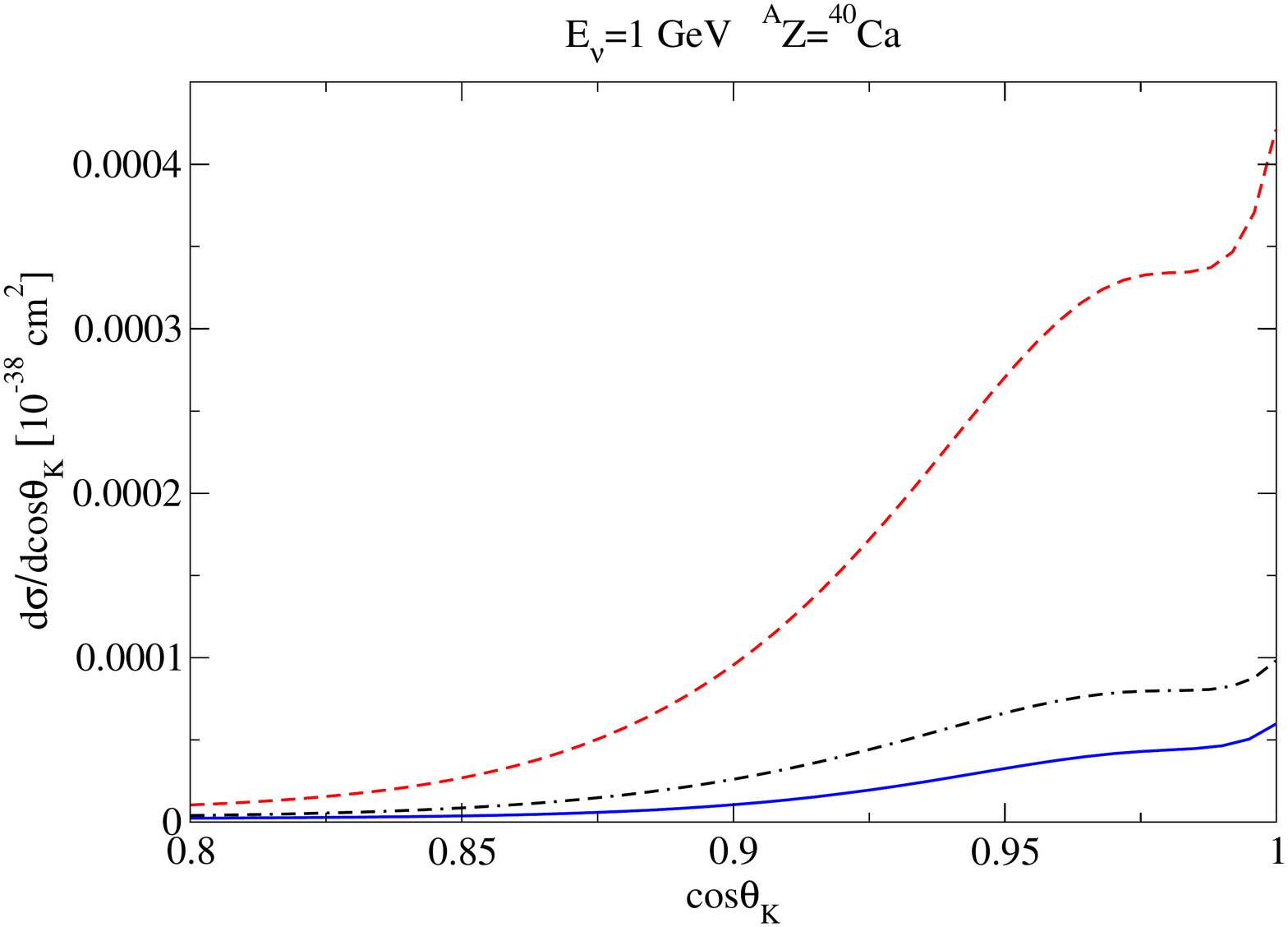}}
\end{center}
\caption{\label{fig:adkk} (Color online) Kaon angular distribution in the Laboratory system at $E_{\bar{\nu}}=1$~GeV for two different nuclei. Lines have the same meaning as in Fig.~\ref{fig:adkmu}.}
\end{figure}

\begin{figure}[h!]
\begin{center}
\makebox[0pt]{\includegraphics[width=0.5\textwidth]{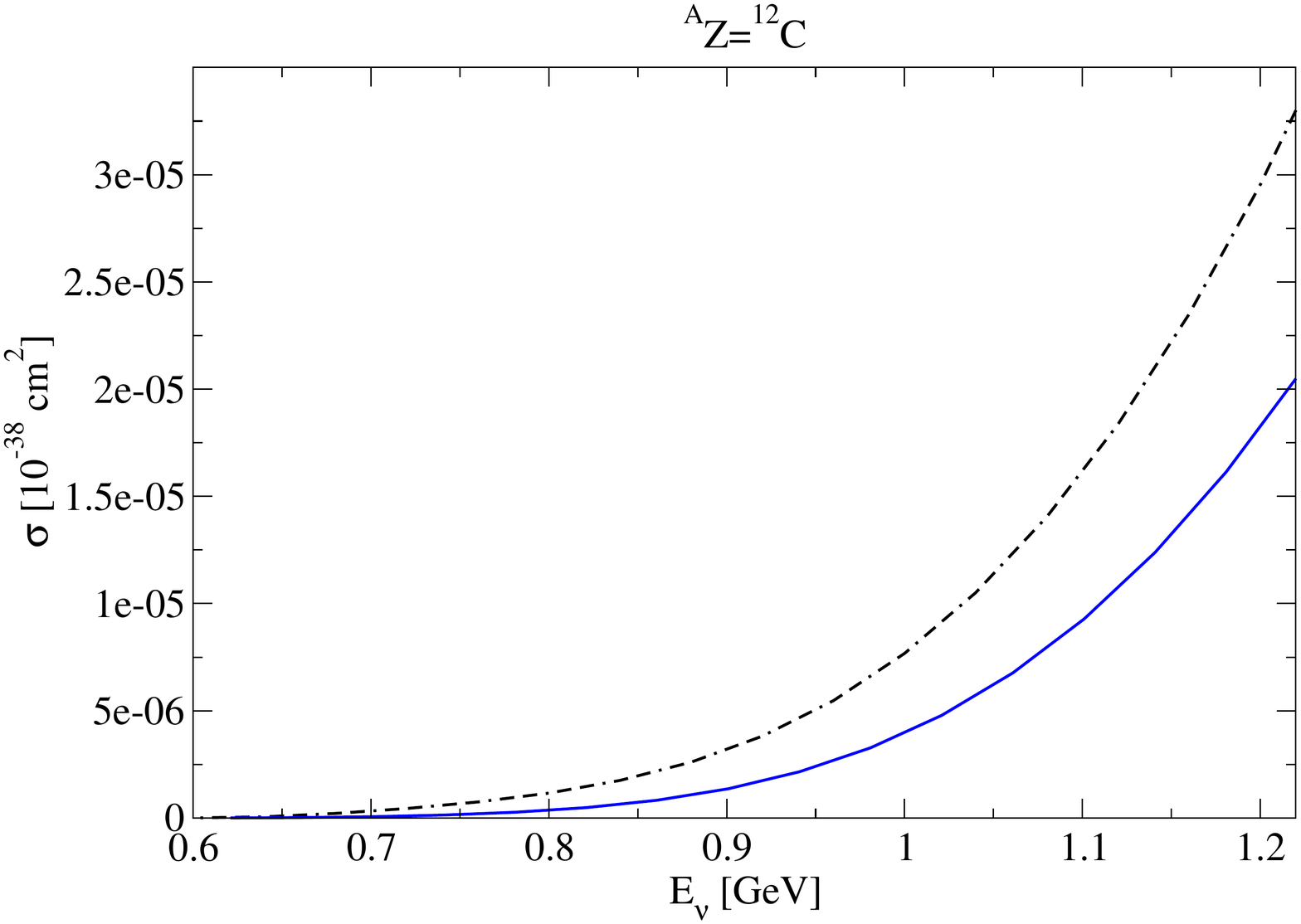}\hspace{0.cm}
              \includegraphics[width=0.5\textwidth]{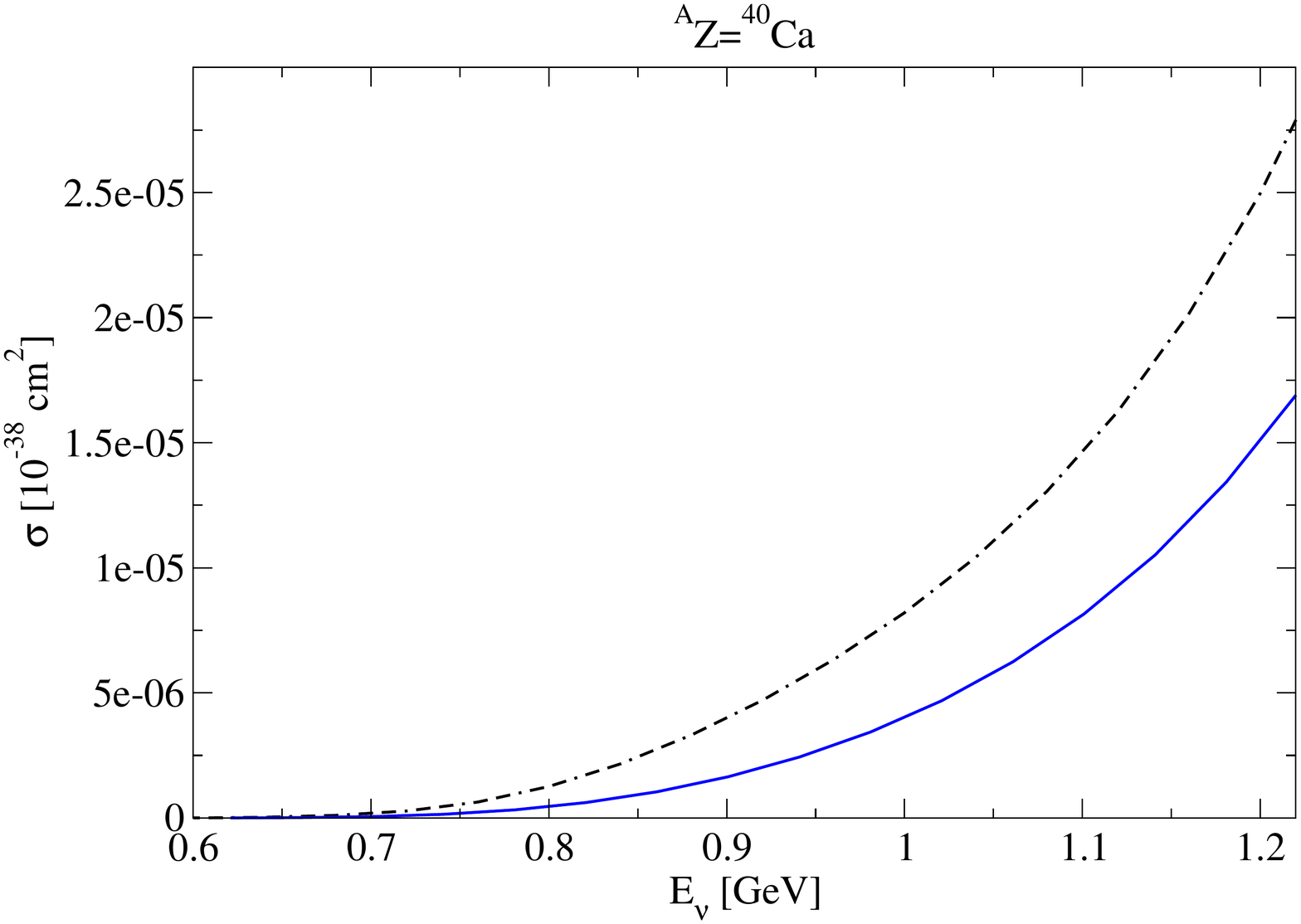}}
\end{center}
\caption{\label{fig:asiged} (Color online) Integrated cross section as a function of the antineutrino energy. Dash-dotted lines are obtained with the full model for kaon plane waves while the solid ones incorporate kaon distortion.}
\end{figure}

The effect of $K^-$ distortion on the energy dependence of the total cross section is shown in Fig.~\ref{fig:asiged}. The energy interval is limited by the validity region of the model for the $\bar{K}$ optical potential, namely $|\vec{p}_K| \leq 1$~GeV$/c$. In presence of the distortion, the cross section is smaller but increases as fast as in the plane-wave case. Nevertheless, one can expect that at higher energies, the absorptive part of the potential becomes more relevant and the cross-section growth slows down, as it happens in the $K^+$ case.   

\begin{figure}[h!]
  \begin{center}
    \includegraphics[scale=0.36]{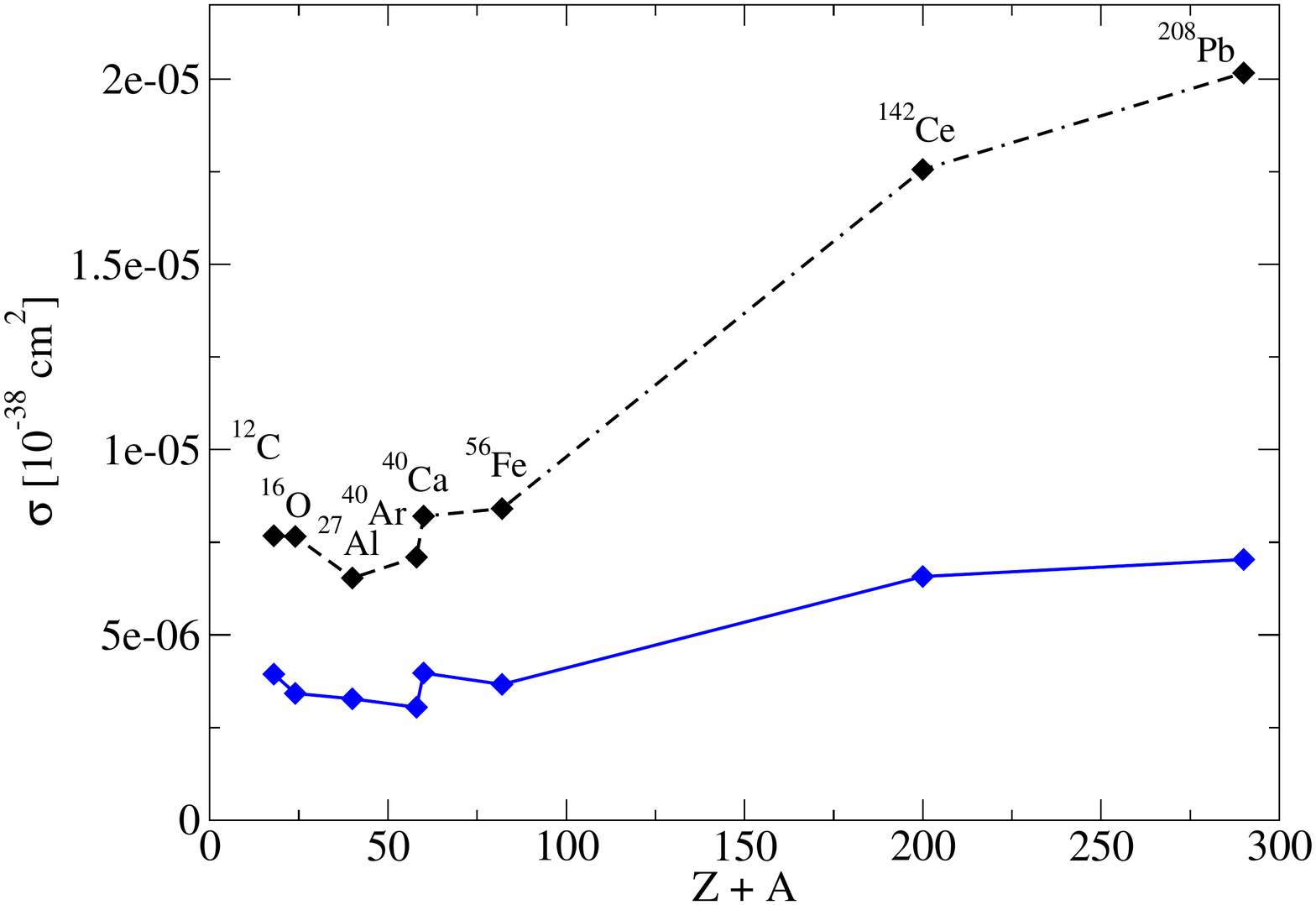}
  \end{center}
  \caption{\label{fig:azrange}  (Color
    online) Total cross section for $\bar{\nu}_{\mu} \, {}^{A}\! Z_{\text{gs}} \to \mu^+ \, {}^{A}\! Z_\text{gs} \, K^-$ 
as a function of $A+Z$ at $E_{\bar{\nu}} =1$~GeV for several nuclei. The dashed (solid) line stands for the calculation without (with) antikaon distortion.}
\end{figure}

Just as for the neutrino-induced reaction, the largest CT current, in absence of distortion scales like $A+Z$ suggesting a quadratic dependence of the cross section on this variable. So we have also studied the cross section dependence on the nuclear target, plotting it as a function of $Z + A$ (Fig.~\ref{fig:azrange}). The comparison with Fig.~\ref{fig:zrange} shows that the cross section are always smaller in the $\bar{\nu}$ case, both without and with kaon distortion. One also observes that the stronger $K^-$ interaction with the medium leads to a flatter $Z + A$ dependence. But apart from these differences, the global trend is very similar for both reactions, which indicates that the role of the nuclear density distributions prevail over the neutrino-nucleon interaction dynamics.  

\section{Summary}
\label{sec:sum}

We have performed state-of-the-art microscopic calculations of weak coherent $K^\pm$ production observables in the few-GeV region. For that we have implemented models for kaon production on nucleons based on chiral SU(3) Lagrangians, supplemented with the excitation of the decuplet-state $\Sigma^*(1385)$ in the $\bar{\nu}$ case. The distortion of the outgoing kaons is treated in a quantum-mechanical way by solving the Klein-Gordon equation with realistic in-medium $K$ and $\bar{K}$ optical potentials. The nuclear density profiles employed are parametrizations of electron scattering data and Hartree-Fock calculations (for the neutrons).  

The resulting cross sections for incident muon neutrinos of 1-2~GeV are small, with cross sections per nucleon much smaller than the corresponding ones on free nucleons. This can be explained by the rather large momentum transferred to the nucleus (due to the large value of the kaon mass compared to the typical kaon momenta) which reduces significantly the nuclear form factors. The situation may be different at higher energies where the present model is not directly applicable. We find similar cross sections for both reactions, with slightly larger values for $\nu$ induced $K^+$ production, even if the dynamics is different. Angular kaon and lepton momentum distributions are forward peaked, as it is normally the case in coherent processes. No significant enhancement for heavy nuclei is observed, in variance with naive expectations. 

In spite of the smallness of the cross sections, our study contributes to  a better and more complete understanding of neutrino interactions with the detector nuclear targets, which is important for current and future neutrino oscillation, proton decay and even dark matter experiments.

\begin{acknowledgments}
  One of us (LAR) thanks Ulrich Mosel for useful comments and suggestions regarding the imaginary part of the kaon optical potential. LAR is grateful to him and to Alexei Larionov for making their parametrization of the kaon-nucleon cross sections available to us. This work has been partially supported by DGI and FEDER funds under contracts FIS2011-28853-C02-01, FIS2011-28853-C02-02 and FIS2011-24149, the Spanish Consolider-Ingenio 2010 Program CPAN (CSD2007-00042), by Generalitat Valenciana under contract PROMETEO/2009/0090, by the EU Hadron-Physics2 project, grant agreement no. 227431, by Joint Action PT2009-0072 and by Junta de Andaluc\'ia grant FQM-225. IRS acknowledges support from the Spanish Ministry of Education and MV from the Japanese Society for the Promotion of Science.
\end{acknowledgments}

\bibliography{kaon}

\end{document}